\documentclass[showpacs,preprintnumbers,amsmath,floatfix,amssymb,12pt]{revtex4}
\usepackage[usenames, dvipsnames]{color}
\usepackage{morefloats}
\usepackage{amssymb}
\usepackage{bm}
\usepackage{graphicx}
\usepackage{epsfig}
\newcommand{\be}{\begin{eqnarray}}
\newcommand{\ee}{\end{eqnarray}}

\begin{document}
\draft
\title{Gamow-Teller Strength distributions in $^{76}\textrm{Ge}$, $^{76,82}\textrm{Se}$, and $^{90,92}\textrm{Zr}$ by the Deformed Proton-neutron QRPA}
\author{Eunja Ha \footnote{ejha@ssu.ac.kr} and Myung-Ki Cheoun \footnote{cheoun@ssu.ac.kr}}
\address{ Department of Physics,
Soongsil University, Seoul 156-743, Korea}
% {\color{red}developed}
% ************************************ red(blue) color is corrected(not corrected).
\begin{abstract}
We developed the deformed proton-neutron quasiparticle random phase
approximation (QRPA) and applied to the evaluation of the Gamow-Teller (GT) transition strength distributions including high-lying excited states, which data becomes recently available beyond one or two nucleon threshold by charge exchange reactions using hundreds of MeV projectiles.
Our calculations started with single-particle states calculated by a deformed axially symmetric Woods-Saxon potential. Neutron-neutron and proton-proton pairing correlations are explicitly taken into account at the deformed Bardeen Cooper Schriffer theory.
Ground state correlations, and two-particle and two-hole mixing states are included in the deformed QRPA. In this work,
we use a realistic two-body interaction given by the Brueckner $G$-matrix based on the CD Bonn potential
to reduce the ambiguity on the nucleon-nucleon interactions inside nuclei. We applied our formalism to the GT
transition strengths for $^{76}$Ge, $^{76,82}$Se, and $^{90,92}\textrm{Zr}$, and compared to available experimental data.
The GT strength distributions were sensitive on the deformation parameter as well as its sign, {\it i.e.}, oblate or prolate.
The Ikeda sum rule,
which is usually thought to be satisfied under the one-body current approximation irrespective of nucleon models, is used to test our numerical calculations and shown to be satisfied without introducing the quenching factor, if high-lying GT excited states are properly taken into account. Most of the GT strength distributions of the nuclei considered in this work turn out to have the high-lying GT excited states beyond one nucleon threshold, which are shown to be consistent with available experimental data.
\end{abstract}

\date{\today}

\maketitle
\section{Introduction}
In the core collapsing supernovae (SNe), medium and heavy elements
are believed to be produced by rapid and slow successive neutron
capture reactions, dubbed as r-process and s-process, respectively. In these processes, many unstable neutron-rich nuclei
are produced iteratively and decay to more stable nuclei at their turning points in the nuclear chart.
These r- and s-processes play vital roles of understanding abundances of the medium and heavy nuclei in the cosmos \cite{Haya04}.

Since most of the nuclei produced in the processes are thought to be more or less deformed, we need to explicitly take into
account the deformation in the nuclear structure and their effects on relevant nuclear reactions in the network calculations of the processes. One interesting process associated with the deformed nuclei may be the rapid proton process (rp-process), which is thought to be occurred on the binary star system composed of a massive compact star and a companion star.
Because of the strong gravitation on the massive star surface, one expects hydrogen rich mass-flow from the companion
star. Since the high density and low temperature on the neutron star crust make electrons degenerated, nuclear beta decays of unstable nuclei may be blocked by the degeneration, while stable nuclei may become unstable with respect to the beta decay. This physical situation gives rise to the nuclear pycno-reactions \cite{Haen91}, where the deformation could be of practical importance on the understanding of the rp-process.

Up to now, many theoretical approaches to understand the nuclear
structure are based on the spherical symmetry \cite{Suhonen14}. In order to describe
the neutron-rich nuclei and their relevant nuclear reactions in the nuclear processes, one needs to
develop theoretical frameworks including explicitly the deformation
\cite{simkovic,saleh,sarriguren_a,sarriguren_b,sarriguren_c,sarriguren_d}. Ref. \cite{simkovic} exploited the Nilsson basis for deformed quasiparticle random phase
approximation (DQRPA). But two-body interactions inside nuclei were derived from the effective separable force.
A realistic two-body interaction derived from the realistic nucleon-nucleon (N-N) force in free space is firstly applied to $2\nu2\beta$ and $0\nu2\beta$ decays within the DQRPA at Ref. \cite{saleh}, where neutron-neutron ($nn$) and proton-proton ($pp$) pairing correlations are considered at the BCS stage.

There are many calculations regarding the deformation effects on the GT strength distributions \cite{sarriguren_a,sarriguren_b,sarriguren_c,sarriguren_d}. Ref. \cite{sarriguren_a} considered the effect with the HF+RPA model using the Skyrme force. These calculations were extended to the deformed QRPA by exploiting the effective separable force \cite{sarriguren_b} or various effective Skyrme forces \cite{sarriguren_c,sarriguren_d}. But, for more {\it ab initio} calculations, it would be more desirable to start from the realistic N-N force in free space and solve the Bethe-Salpeter equation for the N-N interaction in nuclei, {\it i.e.} Brueckner $G$-matrix, as used at Ref. \cite{saleh}.

In this work, we extend our previous spherical QRPA based on the spherical symmetry \cite{Ch93} to the DQRPA \cite{Ha13_1,Ha13_2}.
The spherical QRPA \cite{Ch93} has been exploited as a useful framework for describing the neutrino-induced reactions sensitive
on the nuclear structure of medium-heavy and heavy nuclei \cite{ch10}. For these nuclei, the application of the
shell model may have actual limits because of tremendous increase of mixing configurations as the mass number increases.

This paper is organized as follows.
In Sec. II, we introduce detailed formalism for the DQRPA and the Gamow-Teller (GT) strength.
Numerical results and related discussions are presented in detail at Sec. III. Summary and conclusions are addressed at Sec. IV.
\section{Formalism}
\subsection{Total Hamiltonian}

We start from the following nuclear Hamiltonian
\begin{eqnarray}\label{eq:hamlt}
&&H=H_0 + H_\textrm{int}~,
\\ \nonumber
&&H_{0}= \sum_{ \rho_{\alpha} \alpha \alpha'} \epsilon_{ \rho_{\alpha} \alpha \alpha'}
c_{ \rho_{\alpha} \alpha \alpha'}^{\dagger} c_{ \rho_{\alpha} \alpha \alpha'}~~~~
,
\\ \nonumber
%{\color{blue} &&H_\textrm{int}= \sum_{\alpha \beta \gamma \delta  \rho_{\alpha} \rho_{\beta} \rho_{\gamma}
%\rho_{\delta},~\alpha' \beta' \gamma' \delta'} V_{\alpha \rho_{\alpha}{\alpha'} \beta \rho_{\beta}{\beta'}
%\gamma \rho_{\gamma}{\gamma'} \delta \rho_{\delta}{\delta'}}
%c_{\alpha \rho_{\alpha}\alpha'}^{\dagger}  c_{\beta \rho_{\beta}\beta'}^{\dagger}
%c_{\delta \rho_{\delta}\delta'} c_{\gamma \rho_{\gamma}\gamma'},}
&&H_\textrm{int}= \sum_{ \rho_{\alpha} \rho_{\beta} \rho_{\gamma} \rho_{\delta}, \alpha \beta \gamma \delta,~\alpha' \beta' \gamma' \delta'}
V_{ \rho_{\alpha} \alpha \alpha'  \rho_{\beta} \beta \beta'  \rho_{\gamma} \gamma \gamma'  \rho_{\delta} \delta \delta'}
c_{ \rho_{\alpha} \alpha \alpha'}^{\dagger}  c_{ \rho_{\beta} \beta \beta'}^{\dagger}
c_{ \rho_{\delta} \delta \delta'} c_{ \rho_{\gamma} \gamma \gamma'},
\end{eqnarray}
where the interaction matrix $V$ is the anti-symmetrized interaction with the Baranger Hamiltonian
\cite{Baran63} in which two $-1 \over 2$ factors, from J and T coupling, are included.
Greek letters denote proton or neutron single particle states with a projection $\Omega $ of the total angular momentum on
the nuclear symmetry axis. The projection $\Omega$ is treated as the only good quantum number in the deformed basis.
The $\rho_{\alpha} ~(\rho_{\alpha} =\pm1)$ is a sign of the total angular momentum projection $\Omega$ of the $\alpha$ state. The isospin of
particles is denoted as a Greek letter with prime $(\alpha' , \beta' , \gamma' , \delta')$, while the isospin of quasiparticles is expressed as a Greek letter with double
prime as shown later.

Therefore, the operator $c_{ \rho_{\alpha} \alpha \alpha'}^{\dagger}$
($c_{ \rho_{\alpha} \alpha \alpha'}$) in Eq. (\ref{eq:hamlt}) stands for a usual creation
(destruction) operator of the real particle in a state of $\alpha  \rho_{\alpha}$
with the angular momentum projection $\Omega_{\alpha}$ and the isospin $\alpha'$. Since we assume the time-reversal symmetry, our intrinsic states are twofold-degenerate,
{\it i.e.} $\Omega_{\alpha}$ state and its time-reversed state $-\Omega_{\alpha}$.
$\epsilon_{\rho_{\alpha} \alpha \alpha'}$ means the single particle (s.p.) state energies.

In the cylindrical coordinate, eigenfunctions of a s.p. state and its time-reversed state
in the deformed Woods-Saxon potential are expressed as follows
\begin{eqnarray}\label{eq:Nil}
|\alpha \rho_{\alpha}=+1> =&&\sum_{N n_z} [ b_{N n_z \Omega_{\alpha}}^{(+)}~
|N, n_z, \Lambda_{\alpha}, \Omega_{\alpha}= \Lambda_{\alpha}+ 1/2 > \\ \nonumber
&&+ ~ b_{N n_z \Omega_{\alpha}}^{(-)}|N, n_z, \Lambda_{\alpha}+1, \Omega_{\alpha}= \Lambda_{\alpha}+ 1-1/2 >],\\ \nonumber
|\alpha\rho_{\alpha}=-1> =&&\sum_{N n_z} [b_{N n_z \Omega_{\alpha}}^{(+)}~
|N, n_z, -\Lambda_{\alpha}, \Omega_{\alpha}= -\Lambda_{\alpha}- 1/2 > \\ \nonumber
&&- ~ b_{N n_z \Omega_{\alpha}}^{(-)}~
|N, n_z, -\Lambda_{\alpha}-1, \Omega_{\alpha}= -\Lambda_{\alpha}-1+1/2 >],
\end{eqnarray}
where $N=n_\perp + n_z$ ($n_\perp =2n_\rho + \Lambda $ ) is a
major shell number, and $n_z$ and $n_\rho$ are numbers of nodes of the deformed harmonic oscillator wave function in $z$ and $\rho$ direction, respectively. $\Lambda$ is a projection of orbital angular momentum onto the
nuclear symmetric axis $z$. Coefficients $b^{(+)}_{N n_z \Omega_{\alpha}}$ and $b^{(-)}_{N n_z \Omega_{\alpha}}$
are obtained by the eigenvalue equation of the total Hamiltonian in the Nilsson basis.
The 2nd terms in Eq. (\ref{eq:Nil}) have the same projection $\Omega_{\alpha}$ value as the 1st term, but retain another orbital angular momentum because of a flipped spin. Particle model space is exploited up to $N = 5\hbar\omega$ for deformed basis. In the expansion of the deformed basis to a spherical basis, we considered up to $10 \hbar\omega$ in the spherical basis.

Single particle spectrum obtained by the deformed Woods-Saxon potential is sensitive on the deformation parameter $\beta_2$ defined as
\begin{equation}\label{eq:beta}
R(\theta )= R_0 ( 1 + \beta_2 Y_{20} (\theta ) + \beta_4 Y_{40} (\theta )) ~,
\end{equation}
where $R_0 = 1.2A^{1/3}$fm for the sharp-cut radius $R_0$ \cite{Ring}, and $Y_{20}$ and  $Y_{40}$ are spherical harmonics. The customary parameter, $\epsilon = 3 (\omega_{\perp} - \omega_3) / ( 2 \omega_{\perp} + \omega_3)$, used in the deformed harmonic oscillator is related to as $\beta_2 \approx (2/3)\sqrt{4\pi /5}~ \epsilon $ at the leading order.
In the cylindrical Woods-Saxon potential, we use the following nuclear and spin-orbit potentials \cite{damgaard}
\begin{equation}\label{eq:cws}
V(l )={-V_0 \over {1 + exp(l(\vec{r};r_0, \beta_2, \beta_4} )/a) },~~~~~
V_{so} = -\lambda (\hslash / 2mc)^2 grad V(l)( \vec{\sigma} \times \vec{p}),
\end{equation}
where {\it l} is a distance function of a given point $\vec{r}$ to the nuclear surface represented by Eq. (\ref{eq:beta}).
{\it a} and $\lambda$ are the diffuseness parameter and the strength of spin-orbit potential, respectively.
In this work, we assume $\beta_4 =0$ and use the cylindrical Woods-Saxon potential parameters by Nojarov \cite{Nojarov}.
We transform the Hamiltonian represented by real particles in Eq. (\ref{eq:hamlt}) to the quasiparticle representation
through the Hartree Fock Bogoliubob (HFB) transformation,
\begin{equation}\label{eq:HFB}
a_{\rho_{\alpha} \alpha \alpha''}^{\dagger}=\sum_{\rho_{\beta} \beta \beta'} ( u_{\alpha \alpha'' \beta \beta'}
c_{\rho_{\beta} \beta \beta'}^{\dagger}+v_{\alpha \alpha'' \beta \beta'}c_{\rho_{\beta} \bar{\beta} \beta'}),~
a_{\rho_{\alpha} \bar{\alpha} \alpha''}=\sum_{\rho_{\beta} \beta \beta'}(u_{\bar{\alpha} \alpha'' \bar{\beta} \beta'}
c_{\rho_{\beta} \bar{\beta} \beta'}-v_{\bar{\alpha} \alpha'' \bar{\beta} \beta'}c_{ \rho_{\beta} \beta \beta'}^{\dagger}).
\end{equation}
Since our formalism is intended to include the neutron-proton ($np$) pairing correlations, we denote the isospin of quasiparticles as $\alpha'' ( \beta'') = 1,2$, while the isospin of real particles is denoted as $\alpha (\beta) = p,n$. We assume the time reversal symmetry, which means $ u_{\alpha \alpha'' \beta \beta'} = u^{*}_{{\bar \beta} \alpha'' {\bar \alpha} \beta'} $ and $ v_{\alpha \alpha'' \beta \beta'} = - v^{*}_{{\bar \beta} \alpha'' {\bar \alpha} \beta'} $, and do not allow mixing of different single particle states ($\alpha$ and $\beta$) to the quaisparticle in the deformed state. But, in the spherical state, the quasiparticle states turn out to be mixed with different particle states because each deformed state (basis) is represented by a linear combination of the spherical state (basis). The HFB transformation for each $\rho_{\alpha}$ is then reduced to the following form

\begin{equation} \label{eq:HFB_1}
\left( \begin{array}{c} a_{1}^{\dagger} \\
  a_{2}^{\dagger} \\
  a_{\bar{1}} \\
  a_{\bar{2}}
  \end{array}\right)_{\alpha} =
\left( \begin{array}{cccc}
u_{1p} & u_{1n} & v_{1p} & v_{1n} \\
u_{2p} & u_{2n} & v_{2p} & v_{2n} \\
-v_{1p} & -v_{1n} & u_{1p} & u_{1n} \\
-v_{2p} & -v_{2n} & u_{2p} & u_{2n}
  \end{array}\right)_{\alpha}
\left( \begin{array}{c}
  c_{1}^{\dagger} \\
  c_{2}^{\dagger} \\
  c_{\bar{1}} \\
  c_{\bar{2}}
  \end{array}\right)_{\alpha}
 \end{equation}
and the Hamiltonian is expressed in terms of the quasiparticle as follows
\begin{equation}
H^{'}  = H_{0}^{'} + \sum_{\rho_{\alpha} \alpha \alpha''} E_{\alpha
\alpha'' } a_{\rho_{\alpha} \alpha \alpha''}^{\dagger} a_{\rho_{\alpha} \alpha \alpha''} + H_{qp.int}~.
\end{equation}
Finally, using the transformation of Eq. (\ref{eq:HFB_1}), we obtain the following deformed HFB equation:
\begin{equation} \label{eq:hfbeq}
\left( \begin{array}{cccc} \epsilon_{p}-\lambda_{p} & 0 &
\Delta_{p {\bar p}} & \Delta_{p {\bar n}} \\
0  & \epsilon_{n}-\lambda_{n} & \Delta_{n
{\bar p}} & \Delta_{n {\bar n}} \\
  \Delta_{p {\bar p}} &
 \Delta_{p {\bar n}} & -\epsilon_{p} + \lambda_{p} & 0 \\
  \Delta_{n {\bar p}} &
 \Delta_{n {\bar n}} & 0 & -\epsilon_{n} + \lambda_{n}
  \end{array}\right)_{\alpha}
\left( \begin{array}{c}
u_{\alpha'' p} \\ u_{\alpha'' n} \\ v_{\alpha'' p} \\
v_{\alpha'' n} \end{array}\right)_{\alpha}
 =
 E_{\alpha \alpha''}
\left( \begin{array}{c} u_{\alpha'' p} \\ u_{\alpha'' n} \\
 v_{\alpha'' p} \\
v_{\alpha'' n} \end{array}\right)_{\alpha},
\end{equation}
where $E_{\alpha \alpha''}$ is the energy of a quasiparticle with the isospin quantum number $\alpha''$ in the state $\alpha$.
Pairing potentials in a deformed basis are detailed in the next subsection B. In the present calculation,
we neglect $\Delta_{np}$. This equation therefore reduces to the standard Deformed Hartree Fock Bogoliubov (DHFB) equation \cite{simkovic}.

\subsection{Spherical and deformed wave functions for a single particle state}
Since various mathematical theorems regarding quantum numbers may not be easily used in the deformed basis, it is more convenient to play in the spherical basis. In addition, we exploit the $G$-matrix based on the Bonn potential in order to reduce plausible ambiguities on the N-N interaction inside a deformed nucleus. Since the $G$-matrix is calculated on the spherical basis, we need to represent the $G$-matrix in terms of the deformed basis. Here we present regarding how to transform the deformed wave function to the spherical one.

\begin{figure}
\includegraphics[width=0.8\linewidth]{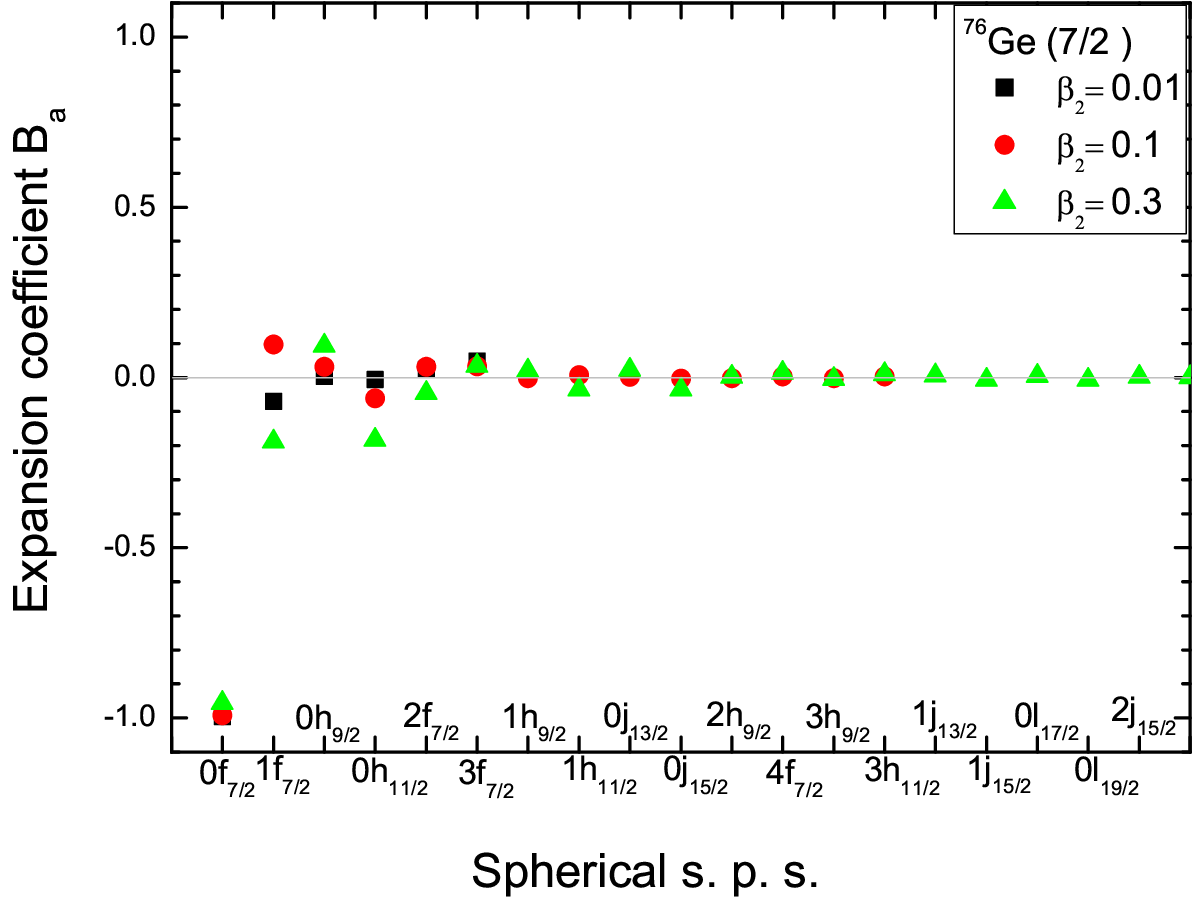}
\caption{(Color online) Dependence of the expansion coefficient $ B_{a}$, Eq. (\ref{eq:sps_ex}), on $\beta_2$ values
in a deformed state $| 7/2 >$ for $^{76}$Ge.}
\label{sps_ex}
\end{figure}

The deformed harmonic oscillator wave function, $|N n_z \Lambda_{\alpha} \Omega_{\alpha} (= \Lambda_{\alpha} + \Sigma )> = |N n_z \Lambda_{\alpha} > |\Sigma>$ in Eq. (\ref{eq:Nil})
can be expanded in terms of the spherical harmonic oscillator wave function $|N_0 l \Lambda_{\alpha} > | \Sigma > $
\begin{eqnarray}\label{eq:trans}
&&|N n_z \Lambda_{\alpha}>~|\Sigma > =
\sum_{N_0 =N, N \pm 2, N \pm 4,...}\sum_{l=N_0, N_0 - 2, N_0 -4,...}
A_{N n_z \Lambda}^{N_0 l, ~n_{r}={N_0 -l \over 2}}~|N_0 l \Lambda_{\alpha}>~|\Sigma >, \\ \nonumber
&& |N_0 l \Lambda_{\alpha}>~|\Sigma > = \sum_{j} C_{l \Lambda_{\alpha}{1 \over 2}\Sigma}^{j \Omega_{\alpha}}
|N_0 l j ~\Omega_{\alpha}>~,
\end{eqnarray}
where the spatial overlap
integral $A_{N n_z \Lambda}^{N_0 l n_r} =<N_0 l \Lambda|N n_z \Lambda
>$ is calculated numerically in the spherical coordinate system.
$ C_{l \Lambda_{\alpha} { 1 \over 2} \Sigma}^{j \Omega_{\alpha}}$ is the
Clebsch-Gordan coefficient of the coupling of the orbital $(l)$ and spin angular momentum (${ 1 \over 2})$ to the total angular momentum $(j)$ with the projection $\Omega_{\alpha}$. Therefore, the expansion of the deformed state $| \alpha \Omega_{\alpha} > = | N n_z \Lambda_{\alpha} \Omega_{\alpha} > $ into the spherical state $ | a \Omega_{\alpha} > = | N_0 l \Lambda_{\alpha} \Sigma >$ can be simply written as
\begin{equation} \label{eq:sps_ex}
|\alpha \Omega_{\alpha}> =\sum_{a} B_{a}^{\alpha}~|a \Omega_{\alpha} > ~, ~B_{a}^{\alpha} = \sum_{N n_z \Sigma} C_{l \Lambda { 1 \over 2} \Sigma}^{j \Omega_{\alpha}}
A_{N n_z \Lambda}^{N_0 l}~b_{N n_z \Sigma} ~,
\end{equation}
where $B_a^{\alpha}$ is the expansion coefficient. Here $a$ indicates quantum numbers ($ N_{0}lj$) of a nucleon state, where major quantum number $N_0$ is related to the radial quantum number $n_r$.
Naturally, the expansion coefficient $B_{a}^{\alpha}$ depends on the deformation parameter $\beta_2$. In order to figure out the dependence, in Fig. \ref{sps_ex}, we show an example of the expansion, where the $B_{a}^{\alpha}$ for $| \alpha > = | 7/2 >$ state is plotted in terms of spherical states (basis) for $\beta_2$ = 0.01, 0.1 and 0.3 cases. The $| \alpha > = | 7/2 >$ state turns out to be composed mainly of $0f_{7/2}, 1f_{7/2}, 0h_{9/2}$ and $0h_{11/2}$ states in the spherical basis.

The pairing potentials in the DHFB Eq. (\ref{eq:hfbeq}) are calculated in the deformed basis by using the $G$-matrix calculated from the realistic Bonn CD potential for the N-N interaction in the following way
\begin{equation} \label{eq:gap}
\Delta_{\alpha p \bar{\alpha}p} = - {1 \over 2}
\sum_{J, c }g_{\textrm{pair}}^{p} F_{\alpha a \bar{\alpha} a}^{J0} F_{\gamma c \bar{\gamma} c}^{J0}
G(aacc,J) (u_{1p_c}^* v_{1p_c} + u_{2p_c}^* v_{2p_c}) ~,
\end{equation}
where $F_{ \alpha a  {\bar \alpha a
}}^{JK}=B_{a}^{\alpha}~B_{a}^{\alpha} ~{(-1)^{j_{a} -\Omega_{\alpha}}}~C^{JK}_{j_{a}
\Omega_{\alpha} j_{a}-\Omega_{\alpha}}(K=\Omega_{\alpha} - \Omega_{\alpha})$ is introduced for the $G$-matrix representation in the deformed basis.
Here $K$, which is a projection number of the total angular momentum $J$ onto the $z$ axis, is selected $K=0$
at the DHFB stage
because we consider pairings of the quasiparticles at $\alpha$ and ${\bar\alpha}$ states. $G(aacc ~J)$ is the two-body (pairwise) scattering matrix in the spherical basis taking into account all possible scattering of nucleon pairs above Fermi surface.

In this work, we include all possible $J$ values which have $K =0$ projection.
$\Delta_{\alpha n \bar{\alpha}n}$ is the same as Eq. (\ref{eq:gap}) with replacement of $n$ by $p$.
In order to renormalize the $G$-matrix, strength parameters,
$g_{\textrm{pair}}^{p}$ and $g_{\textrm{pair}}^{n}$ are multiplied to the $G$-matrix \cite{Ch93} by adjusting the pairing potentials to
the empirical pairing potentials, $\Delta_{p}^{emp}$ and $\Delta_{n}^{emp}$.
The empirical pairing potentials of protons and neutrons are evaluated by the following symmetric five term formula for neighboring nuclei
 \begin{eqnarray} \label{eq:gap_p}
 {\Delta_p}^\textrm{emp} &=& {1 \over 8} [ M(Z+2, N)- 4M(Z+1, N)+ 6M(Z, N)\\
 & &-4M(Z-1, N)+ M(Z-2, N) ]~, \nonumber
 \end{eqnarray}
 \begin{eqnarray} \label{eq:gap_n}
  {\Delta_n}^\textrm{emp} &=& {1 \over 8} [ M(Z, N+2)- 4M(Z, N+1)+ 6M(Z, N) \\
 & &-4M(Z, N-1)+ M(Z, N-2) ] ~, \nonumber
 \end{eqnarray}
%
% \begin{eqnarray}
%  {\Delta_{pn}}^\textrm{emp} &=& \pm {1 \over 4} \{ 2[ M(Z, N+1)+ M(Z, N-1)+ M(Z-1, N)+ M(Z+1, N)] \\
% & & - [ M(Z+1, N+1)+ M(Z-1, N+1)+M(Z-1, N-1)+ M(Z+1, N-1)]  \nonumber \\
% & & - 4M(Z,N) \}~,  \nonumber
% \label{eq:gap-pn}
% \end{eqnarray}
where signs in the +(--) stand for even(odd) mass nuclei.
As for masses in Eqs. (\ref{eq:gap_p}) and (\ref{eq:gap_n}), we use empirical masses.

\subsection{Description of an excited state by the DQRPA}

We take the ground state of an even-even target nucleus as the DBCS vacuum for a quasiparticle. In the following,
we show how to generate an excited state in a deformed nucleus.
Since deformed nuclei have two different frames, laboratory and intrinsic frames,
we need to consider a relationship of the two frames.
The GT excited state in the intrinsic frame of even-even nuclei, which is described by operating a phonon operator to the QRPA vacuum ${\cal Q}_{m,K}^{\dagger} | QRPA>$, can be transformed to the wave function in the laboratory frame by using the Wigner function ${\cal D}^{1}_{MK} ( \phi, \theta, \psi) $ as follows
\begin{eqnarray}
| 1M(K),m>&=&\sqrt{ 3 \over 8 \pi^2} ~
 {\cal D}_{M K}^1 (\phi, \theta, \psi){\cal Q}_{m,K}^{\dagger} | QRPA>
 ~~(\textrm{for}~ K=0),
 \\ \nonumber
 | 1M(K),m>&=&\sqrt{ 3 \over 16 \pi^2}~[
 {\cal D}_{M K}^1 (\phi, \theta, \psi){\cal Q}_{m,K}^{\dagger}
\\ \nonumber & &+(-1)^{1+K}{\cal D}_{M -K}^1 (\phi, \theta, \psi){\cal Q}_{m,-K}^{\dagger} ] | QRPA>
~~(\textrm{for}~ K=\pm 1).
\end{eqnarray}
Here $| QRPA>$ is the correlated ground state in the intrinsic frame, and $| 1M(K),m>$ is the proton-neutron DQRPA wave function for the Gamow-Teller excited state in the laboratory frame.
$M$($K$) is a projection of the total angular momentum onto the $z$ (the nuclear symmetry) axis, where the $K$ is accepted as only a good quantum number in deformed nuclei.
The DQRPA phonon creation operator ${\cal Q}^{\dagger}_{m,K}$ acting on the ground state is given as
\begin{equation}
{\cal Q}^{\dagger}_{m,K}  =\sum_{\rho_{\alpha} \alpha \alpha'' \rho_{\beta} \beta \beta''}
[ X^{m}_{( \alpha \alpha'' \beta \beta''K)} A^{\dagger}( \alpha \alpha'' \beta \beta'' K)
- Y^{m}_{( \alpha \alpha'' \beta \beta''K)} {\tilde A}( \alpha \alpha'' \beta \beta'' K)],
\end{equation}
with pairing creation and annihilation operators composed by two quasiparticles defined as
\begin{equation}
 A^{\dagger}( \alpha \alpha'' \beta \beta'' K)  =
 a^{\dagger}_{ \alpha \alpha''} a^{\dagger}_{\beta \beta''},
~~~{\tilde A}( \alpha \alpha'' \beta \beta'' K)  =
 a_{\beta \beta''} a_{\alpha \alpha''}.
\end{equation}
The quasiparticle pairs in two-particle states, $\alpha$ and $ \beta$ with the parity $\pi_{\alpha (\beta)}$, are chosen by the selection rules
\begin{eqnarray}
 K=0 ~~~ &&\Omega_{\alpha}-\Omega_{\beta}=0, ~~ \pi_{\alpha} \pi_{\beta} = 1 \\
 K=1 ~~~ &&
  \begin{cases}
     \Omega_{\alpha}-\Omega_{\beta}=1, &\pi_{\alpha} \pi_{\beta} = 1  \\ \nonumber
   - \Omega_{\alpha}+\Omega_{\beta}=1, &\pi_{\alpha} \pi_{\beta} = 1 \\
     \Omega_{\alpha}+\Omega_{\beta}=1, &\pi_{\alpha} \pi_{\beta} = 1.
   \end{cases}
\end{eqnarray}
We note that our our quasiparticle pairs include all combinations of particle states and their time reversed states and $K = -1$ and $K =1$ modes are degenerated through the time reversal symmetry. Two-body wave functions in the deformed basis are calculated from the spherical basis as follows
\begin{eqnarray}
|\alpha {\bar \beta}> &=&\sum_{ab J} F_{ \alpha a  {\bar \beta} b}^{JK}|ab ,JK >, \\  \nonumber
|{\bar \alpha} \beta > &=&\sum_{ab J} F_{ {\bar \alpha} a  \beta b}^{JK}|ab ,JK >, \\ \nonumber
|\alpha \beta> &=&\sum_{ab J} F_{ \alpha a  \beta b}^{JK}|ab ,JK >,
\end{eqnarray}
where two body wave function in the spherical basis, $|ab ,JK >$, and transformation coefficient are given as
\begin{equation}
|ab,~JK >=\sum_{J} C^{JK}_{j_a \Omega_a j_b \Omega_b}
|a \Omega_a>|b \Omega_b> ~,~ F_{ {\bar \alpha} a  \beta b }^{JK}=B_{a}^{\alpha}~B_{b}^{\beta}
(-1)^{j_{a} -\Omega_{\alpha}}~C^{JK}_{j_{a} -\Omega_{\alpha} j_{b}
 \Omega_{\beta}} ~,
\end{equation}
where the phase factor $(-1)^{j_{a -\Omega_{\alpha}}}$
comes from the time-reversed state ${\bar\alpha}$. The expansion coefficient for a single particle state, $B_a^{\alpha}$, is defined in Eq. (\ref{eq:sps_ex}).

\subsection{Deformed QRPA equation}
By taking the same approach as the derivation of the QRPA equation in Ref. \cite{Suhonen}, we obtain the Deformed QRPA (DQRPA) equation. But, in this paper, we present more general formalism, which includes the $np$ pairing correlations, for further study. They become important for the description neutron deficient (proton-rich) nuclei or light nuclei. Our DQRPA equation is finally given as in the deformed basis

\begin{eqnarray}\label{qrpaeq}
&&\left( \begin{array}{cccccc}
    ~A_{\alpha \beta \gamma \delta}^{1111}(K) &~ A_{\alpha \beta \gamma \delta}^{1122}(K) &
    ~ A_{\alpha \beta \gamma \delta}^{1112}(K) &
    ~ B_{\alpha \beta \gamma \delta}^{1111}(K) &~ B_{\alpha \beta \gamma \delta}^{1122}(K) &
    ~ B_{\alpha \beta \gamma \delta}^{1112}(K) \\
     ~A_{\alpha \beta \gamma \delta}^{2211}(K) &~ A_{\alpha \beta \gamma \delta}^{2222}(K) &
     ~A_{\alpha \beta \gamma \delta}^{2212}(K) &
           ~B_{\alpha \beta \gamma \delta}^{2211}(K) & ~B_{\alpha \beta \gamma \delta}^{2222}(K) &
           ~B_{\alpha \beta \gamma \delta}^{2212}(K) \\
           ~A_{\alpha \beta \gamma \delta}^{1211}(K) & ~A_{\alpha \beta \gamma \delta}^{1222}(K) &
           ~A_{\alpha \beta \gamma \delta}^{1212}(K) &
           ~B_{\alpha \beta \gamma \delta}^{1211}(K) & ~B_{\alpha \beta \gamma \delta}^{1222}(K) &
           ~B_{\alpha \beta \gamma \delta}^{1212}(K) \\
                            &                  &            &
                            &                  &            \\
           - B_{\alpha \beta \gamma \delta}^{1111}(K) & -B_{\alpha \beta \gamma \delta}^{1122}(K) &
            -B_{\alpha \beta \gamma \delta}^{1112}(K) &
           - A_{\alpha \beta \gamma \delta}^{1111}(K) & -A_{\alpha \beta \gamma \delta}^{1122}(K) &
           -A_{\alpha \beta \gamma \delta}^{1112}(K) \\
           - B_{\alpha \beta \gamma \delta}^{2211}(K) & -B_{\alpha \beta \gamma \delta}^{2222}(K) &
           -B_{\alpha \beta \gamma \delta}^{2212}(K) &
           - A_{\alpha \beta \gamma \delta}^{2211}(K) & -A_{\alpha \beta \gamma \delta}^{2222}(K) &
           -A_{\alpha \beta \gamma \delta}^{2212}(K) \\
           - B_{\alpha \beta \gamma \delta}^{1211}(K) & -B_{\alpha \beta \gamma \delta}^{1222}(K) &
           -B_{\alpha \beta \gamma \delta}^{1212}(K) &
           - A_{\alpha \beta \gamma \delta}^{1211}(K) & -A_{\alpha \beta \gamma \delta}^{1222}(K) &
           -A_{\alpha \beta \gamma \delta}^{1212}(K) \end{array} \right)\\ \nonumber ~~&& \times
\left( \begin{array}{c}   {\tilde X}_{(\gamma 1 \delta 1)K}^{m}  \\ {\tilde X}_{(\gamma 2 \delta 2)K}^{m} \\
  {\tilde X}_{(\gamma 1 \delta 2)K}^{m} \\
     \cr {\tilde Y}_{(\gamma 1 \delta 1)K}^{m} \\ {\tilde Y}_{(\gamma 2 \delta 2)K}^{m} \\
     {\tilde Y}_{(\gamma 1 \delta 2)K}^{m}  \end{array} \right)
 = \hbar {\Omega}_K^{m}
 \left ( \begin{array}{c} {\tilde X}_{(\alpha 1 \beta 1)K}^{m}  \\
{\tilde X}_{(\alpha 2 \beta 2)K}^{m} \\
 {\tilde X}_{(\alpha 1 \beta 2)K}^{m} \\  \\
{\tilde Y}_{(\alpha 1 \beta 1)K}^{m} \\ {\tilde Y}_{(\alpha 2 \beta 2)K}^{m} \\ {\tilde
Y}_{(\alpha 1 \beta 2)K}^{m}  \end{array} \right)  ~,
\end{eqnarray}
where 1 and 2 denote isospins of quasiparticles {\it i.e.} quasiprotons and quasineutrons as denoted $\alpha'' (\beta'')$ in previous sections.
The amplitudes $X^m_{(\alpha \alpha''  \beta \beta'')K }$ and $Y^m_{(\alpha \alpha''  \beta \beta'')K}$,
which stand for forward and backward going
amplitudes from state ${ \alpha \alpha'' }$ to ${\beta  \beta''}$, are related to
$\tilde{X^m}_{(\alpha \alpha'' \beta \beta'')K}=\sqrt2 \sigma_{\alpha \alpha'' \beta \beta''} X^m_{(\alpha \alpha''
 \beta \beta'')K}$
and $\tilde{Y^m}_{(\alpha \alpha'' \beta \beta'')K}=\sqrt2 \sigma_{\alpha \alpha'' \beta \beta''}
Y^m_{(\alpha \alpha'' \beta \beta'')K}$ in Eq. (\ref{qrpaeq}), where $\sigma_{\alpha \alpha'' \beta \beta''}$ = 1 if $\alpha = \beta$ and $\alpha''$ =
$\beta''$, otherwise $\sigma_{\alpha \alpha'' \beta \beta'' }$ = $\sqrt 2$ \cite{Ch93}. If we neglect the $np$ pairing, {\it i.e.} take only 1212 terms and $\alpha \beta \gamma \delta = (p n p' n')$ in the matrices, Eq. (\ref{qrpaeq}) becomes the proton-neutron DQRPA at Ref. \cite{saleh}.

The A and B matrices in Eq. (\ref{qrpaeq}) are given by
\begin{eqnarray} \label{eq:mat_A}
A_{\alpha \beta \gamma \delta}^{\alpha'' \beta'' \gamma'' \delta''}(K)  = && (E_{\alpha
   \alpha''} + E_{\beta \beta''}) \delta_{\alpha \gamma} \delta_{\alpha'' \gamma''}
   \delta_{\beta \delta} \delta_{\beta'' \delta''}
       - \sigma_{\alpha \alpha'' \beta \beta''}\sigma_{\gamma \gamma'' \delta \delta''}\\ \nonumber
   &\times&
   \sum_{\alpha' \beta' \gamma' \delta'}
   [-g_{pp} (u_{\alpha \alpha''\alpha'} u_{\beta \beta''\beta'} u_{\gamma \gamma''\gamma'} u_{\delta \delta''\delta'}
   +v_{\alpha \alpha''\alpha'} v_{\beta \beta''\beta'} v_{\gamma \gamma''\gamma'} v_{\delta \delta''\delta'} )
    ~V_{\alpha \alpha' \beta \beta',~\gamma \gamma' \delta \delta'}
    \\ \nonumber  &-& g_{ph} (u_{\alpha \alpha''\alpha'} v_{\beta \beta''\beta'}u_{\gamma \gamma''\gamma'}
     v_{\delta \delta''\delta'}
    +v_{\alpha \alpha''\alpha'} u_{\beta \beta''\beta'}v_{\gamma \gamma''\gamma'} u_{\delta \delta''\delta'})
    ~V_{\alpha \alpha' \delta \delta',~\gamma \gamma' \beta \beta'}
     \\ \nonumber  &-& g_{ph} (u_{\alpha \alpha''\alpha'} v_{\beta \beta''\beta'}v_{\gamma \gamma''\gamma'}
     u_{\delta \delta''\delta'}
     +v_{\alpha \alpha''\alpha'} u_{\beta \beta''\beta'}u_{\gamma \gamma''\gamma'} v_{\delta \delta''\delta'})
    ~V_{\alpha \alpha' \gamma \gamma',~\delta \delta' \beta \beta' }],
\end{eqnarray}
\begin{eqnarray} \label{eq:mat_B}
B_{\alpha \beta \gamma \delta}^{\alpha'' \beta'' \gamma'' \delta''}(K)  =
 &-& \sigma_{\alpha \alpha'' \beta \beta''} \sigma_{\gamma \gamma'' \delta \delta''}
  \\ \nonumber &\times&
 \sum_{\alpha' \beta' \gamma' \delta'}
  [g_{pp}
  (u_{\alpha \alpha''\alpha'} u_{\beta \beta''\beta'}v_{\gamma \gamma''\gamma'} v_{\delta \delta''\delta'}
   +v_{\alpha \alpha''\alpha'} v_{{\bar\beta} \beta''\beta'}u_{\gamma \gamma''\gamma'} u_{{\bar\delta} \delta''\delta'} )
   ~ V_{\alpha \alpha' \beta \beta',~\gamma \gamma' \delta \delta'}\\ \nonumber
     &- & g_{ph} (u_{\alpha \alpha''\alpha'} v_{\beta \beta''\beta'}v_{\gamma \gamma''\gamma'}
     u_{\delta \delta''\delta'}
    +v_{\alpha \alpha''\alpha'} u_{\beta \beta''\beta'}u_{\gamma \gamma''\gamma'} v_{\delta \delta''\delta'})
   ~ V_{\alpha \alpha' \delta \delta',~\gamma \gamma' \beta \beta'}
     \\ \nonumber  &- & g_{ph} (u_{\alpha \alpha''\alpha'} v_{\beta \beta''\beta'}u_{\gamma \gamma''\gamma'}
      v_{\delta \delta''\delta'}
     +v_{\alpha \alpha''\alpha'} u_{\beta \beta''\beta'}v_{\gamma \gamma''\gamma'} u_{\delta \delta''\delta'})
   ~ V_{\alpha \alpha' \gamma \gamma',~\delta \delta' \beta \beta'}],
\end{eqnarray}
where $u$ and $v$
coefficients are determined from DHFB calculation with the
pairing strength parameters $g_{\textrm{pair}}^{n} , g_{\textrm{pair}}^{p}$ and $g_{\textrm{pair}}^{np}$ adjusted to the
empirical pairing gaps $\Delta_{nn} , \Delta_{pp} $ and
$\Delta_{np}$, respectively. $E_{\alpha \alpha''}$ indicates the
quasiparticle energy of the state $\alpha$ with the
quasiparticle isospin $\alpha''$. The two body interactions $V_{\alpha \beta,~\gamma \delta}$ and $V_{\alpha \delta,~\gamma \beta}$
are particle-particle
and particle-hole matrix elements of the residual N-N interaction $V$, respectively, in the deformed state. They are calculated from the $G$-matrix in the spherical basis as follows
\begin{eqnarray}
V_{\alpha \alpha' \beta \beta',~\gamma \gamma' \delta \delta'}&= &
-\sum_{J} \sum_{abcd }
 F_{ \alpha a  \beta b }^{JK} F_{\gamma c \delta d}^{JK}G(a\alpha'b\beta'c\gamma'd\delta',~ J) ~,
\\ \nonumber
V_{\alpha \alpha' \delta \delta',~\gamma \gamma' \beta \beta'}&=&
 \sum_{J} \sum_{abcd }
 F_{ \alpha a \delta d}^{JK'} F_{\gamma c \beta b}^{JK'}G(a\alpha'd\delta'c\gamma'b\beta',~ J) ~,
\\ \nonumber
V_{\alpha \alpha' \gamma \gamma',~\delta \delta' \beta \beta'}&=&
\sum_{J} \sum_{abcd }
 F_{ \alpha a  \gamma c}^{JK} F_{\beta b \delta d}^{JK} G(a\alpha'c\gamma'd\delta'b\beta',~ J)~,
\end{eqnarray}
where $F_{ \alpha a \beta b
}^{JK'}=B_{a}^{\alpha}~B_{b}^{\beta} ~C^{JK'}_{j_{a}
\Omega_{\alpha} j_{b}\Omega_{\beta}}$
with $K'=\Omega_{\alpha}+\Omega_{\beta}$. The $G$-matrices are two
body particle - particle matrix elements obtained in spherical basis as
solutions of the Bethe - Goldstone equation from the Bonn potential \cite{Ho81}.
If we do not consider the $np$ pairing correlations, the A and B matrices
can be expressed in the following simple form, which are the same as the those of Ref. \cite{saleh},
\begin{eqnarray}\label{eq:rpa_a}
A_{\alpha \beta \gamma \delta}^{p n p' n'}(K)  = && (E_{\alpha
   p} + E_{\beta n}) \delta_{\alpha \gamma} \delta_{p p'}
   \delta_{\beta \delta} \delta_{n n'}\\ \nonumber
      &+ &2~  [g_{pp} (u_{\alpha p} u_{\beta n} u_{\gamma p'} u_{\delta n'}
   +v_{\alpha p} v_{\beta n} v_{\gamma p'} v_{\delta n'} )
    ~V_{\alpha \beta,~\gamma \delta}
    \\ \nonumber  &+ & g_{ph} (u_{\alpha p} v_{\beta n}u_{\gamma p'}
     v_{\delta n'}
    +v_{\alpha p} u_{\beta n}v_{\gamma p'} u_{\delta n'})
    ~V_{\alpha \delta,~\gamma \beta }],
\end{eqnarray}
\begin{eqnarray}\label{eq:rpa_b}
B_{\alpha \beta \gamma \delta}^{p n p' n'}(K)  =
 &-&2~ [g_{pp}
  (u_{\alpha p} u_{\beta n}v_{\gamma p'} v_{\delta n'}
   +v_{\alpha p} v_{\beta n}u_{\gamma p'} u_{\delta n'} )
   ~ V_{\alpha \beta,~\gamma \delta}\\ \nonumber
     &- & g_{ph} (u_{\alpha p} v_{\beta n}v_{\gamma p'}
     u_{\delta n'}
    +v_{\alpha p} u_{\beta n}u_{\gamma p'} v_{\delta n'})
   ~ V_{\alpha \delta,~\gamma \beta }] ~,
\end{eqnarray}
where the last terms of Eqs. (\ref{eq:mat_A}) and (\ref{eq:mat_B}) are zero
for the proton-neutron DQRPA because of the charge conservation.

\subsection{Description of Gamow-Teller Transition}
The ${\beta}^{\pm}$ decay operator, ${\hat\beta }_{1 \mu}^{\pm}$, is defined in the intrinsic frame as follows
\begin{equation} \label{eq:btop}
{\hat\beta }_{1 \mu}^{-}  = \sum_{\alpha \beta }
< \alpha p |\tau^{+} \sigma_K | \beta n >  c_{\alpha p}^{\dagger} {\tilde c}_{\beta n},~
{\hat\beta}_{1 \mu}^{+}  =  {( { \hat\beta}_{1 \mu}^{-} )}^\dagger  = {(-)}^{\mu}
{ \hat\beta}_{1,-\mu}^{-},
\end{equation}
in which the ${\hat \beta}_{1 \mu }^{\pm}$ transition operators are related with those in the laboratory system
$ {\hat \beta}_{M }^{\pm}$ operator as follows
\begin{equation}
{\hat\beta}_{M }^{\pm} =\sum_{\mu} {\cal D}_{M \mu}^1 (\phi, \theta, \psi) \hat\beta_{1 \mu }^{\pm}.
\end{equation}
The $\beta^{\pm}$ transition amplitudes from the ground state of an initial (parent) nucleus
to the excited state of a daughter nucleus, {\it i.e.} the one phonon state
$K^+$ in a final nucleus, are written as
\begin{eqnarray}
&&< K^+, m | {\hat\beta}_{K }^- | ~QRPA >  \\ \nonumber
&&= \sum_{\alpha \alpha''\rho_{\alpha} \beta \beta''\rho_{\beta}}{\cal N}_{\alpha \alpha''\rho_{\alpha}
 \beta \beta''\rho_{\beta} }
 < \alpha \alpha''p \rho_{\alpha}|  \sigma_K | \beta \beta''n \rho_{\beta}>
 [ u_{\alpha \alpha'' p} v_{\beta \beta'' n} X_{(\alpha \alpha''\beta \beta'')K}^{m} +
v_{\alpha \alpha'' p} u_{\beta \beta'' n} Y_{(\alpha \alpha'' \beta \beta'')K}^{m}], \\ \nonumber
&&< K^+, m | {\hat \beta}_{K }^+ | ~QRPA >  \\ \nonumber
&&= \sum_{\alpha \alpha'' \rho_{\alpha} \beta \beta''\rho_{\beta}}{\cal N}_{\alpha \alpha'' \beta \beta'' }
 < \alpha \alpha''p \rho_{\alpha}|  \sigma_K | \beta \beta''n \rho_{\beta}>
 [ u_{\alpha \alpha'' p} v_{\beta \beta'' n} Y_{(\alpha \alpha'' \beta \beta'')K}^{m} +
v_{\alpha \alpha'' p} u_{\beta \beta'' n} X_{(\alpha \alpha'' \beta \beta'')K}^{m} ]
\end{eqnarray}
where $|~QRPA >$ denotes the correlated QRPA ground state in the intrinsic frame and
the nomalization factor is given as $ {\cal N}_{\alpha \alpha'' \beta
 \beta''} (J) = \sqrt{ 1 - \delta_{\alpha \beta} \delta_{\alpha'' \beta''} (-1)^{J + T} }/
 (1 + \delta_{\alpha \beta} \delta_{\alpha'' \beta''}). $ The Wigner functions are disappeared by using the orthogonality of two Wigner functions from the operator and the excited state, respectively. This form is also easily reduced to the results by proton-neutron DQRPA without the $np$ pairing
\begin{eqnarray} \label{eq:tram_d}
&&< K^+, m | {\hat \beta}_{K }^- | ~QRPA > \\ \nonumber
&&= {\sum_{\alpha \rho_{\alpha} \beta \rho_{\beta}}} < \alpha p \rho_{\alpha} |\tau^{+} \sigma_K
| \beta n \rho_{\beta}> ~[ u_{\alpha p} v_{ \beta n} X_{(\alpha p \beta n)K}^{m} + v_{\alpha p} u_{\beta n } Y_{(\alpha p \beta n)K}^{m} ] ~,\\ \nonumber
&&< K^+, m | {\hat \beta}_{K }^+ | ~QRPA > ~\\ \nonumber
&&= {\sum_{\alpha \rho_{\alpha} \beta \rho_{\beta}}} < \alpha p \rho_{\alpha} |\tau^{+} \sigma_K
| \beta n \rho_{\beta}> ~[ u_{\alpha p} v_{ \beta n} Y_{(\alpha p \beta n)K}^{m} + v_{\alpha p
} u_{\beta n } X_{(\alpha p \beta n)K}^{m} ].
\end{eqnarray}
In this work we calculate the transition amplitudes by two different ways. One is to directly calculate in a deformed basis and the other is to calculate them in a spherical basis by using the expansion of the deformed basis in Eq. (\ref{eq:sps_ex}). In particular, the latter method can be widely applied to any types' transition operators. In the deformed basis, the single particle matrix elements of ${\langle \alpha_p \rho_\alpha \vert \tau^{+} \sigma_K  \vert \beta_n \rho_\beta \rangle}$ can be expressed \cite{simkovic},
\begin{equation}\label{eq:tram_d1}
< \alpha p \rho_\alpha \vert \tau^{+} \sigma_{K=0}  \vert \beta n \rho_\beta >
= \delta_{\Omega_p \Omega_n }\rho_\alpha \sum_{N n_z }[  b_{N n_z \Omega_p}^{(+)} b_{N n_z \Omega_n}^{(+)}
 -b_{N n_z \Omega_p}^{(-)} b_{N n_z \Omega_n}^{(-)}],
\end{equation}
\begin{eqnarray} \label{eq:tram_d2}
< \alpha p \rho_\alpha \vert \tau^{+} \sigma_{K=1}  \vert \beta n \rho_\beta >
&&= -\sqrt2 \delta_{\Omega_p {\Omega_n}+1 } \sum_{N n_z } b_{N n_z \Omega_p}^{(+)} b_{N n_z \Omega_n}^{(-)}
~~(\rho_\alpha =\rho_\beta =+1)\\ \nonumber
&&= +\sqrt2 \delta_{\Omega_p {\Omega_n}+1 } \sum_{N n_z } b_{N n_z \Omega_p}^{(-)} b_{N n_z \Omega_n}^{(+)}
~~ ( \rho_\alpha =\rho_\beta =-1 )\\ \nonumber
&&= -\sqrt2 \delta_{\Omega_p {1 \over 2}} \delta_{\Omega_n -{1 \over 2}} \sum_{N n_z } b_{N n_z \Omega_p}^{(+)} b_{N n_z \Omega_n}^{(+)}
 ~~( \rho_\alpha =+1, \rho_\beta =-1 ),
\end{eqnarray}
\begin{eqnarray} \label{eq:tram_d3}
< \alpha p \rho_\alpha \vert \tau^{+} \sigma_{K=-1} \vert \beta n \rho_\beta >
&&= \sqrt2 \delta_{\Omega_p {\Omega_n}-1 } \sum_{N n_z } b_{N n_z \Omega_p}^{(-)} b_{N n_z \Omega_n}^{(+)}
~~( \rho_\alpha =\rho_\beta =+1 )  \\ \nonumber
&&= -\sqrt2 \delta_{\Omega_p {\Omega_n}-1 }\sum_{N n_z } b_{N n_z \Omega_p}^{(+)} b_{N n_z \Omega_n}^{(-)}
~~( \rho_\alpha =\rho_\beta =-1 ) \\ \nonumber
&&= +\sqrt2 \delta_{\Omega_p -{1 \over 2}} \delta_{\Omega_n {1 \over 2}}\sum_{N n_z } b_{N n_z \Omega_p}^{(+)} b_{N n_z \Omega_n}^{(+)}
 ~~( \rho_\alpha =+1, \rho_\beta =-1 ).
\end{eqnarray}
 On the other hand the single particle matrix elements, ${\langle \alpha_p \rho_\alpha \vert \tau^{+} \sigma_K  \vert \beta_n \rho_\beta \rangle}$, can be written in spherical basis
\begin{eqnarray}\label{babo}
< \alpha p \rho_\alpha \vert \tau^{+} \sigma_{K} \vert \beta n \rho_\beta >
&&= {\sum_{ab}} F_{{\alpha_p}a {\beta_n}b }^{1K} {< a p||\tau^{+} \sigma_K || b n > \over \sqrt3 }, \\
< a p||\tau^{+} \sigma_K || b n >
&&=  \sqrt6 \delta_{n_a n_b} \delta_{l_a l_b} \sqrt{2j_{a} +1} \sqrt{2j_{b} +1} (-1)^{l_{a} + j_{a} + { 3 \over 2}}
\left \{ \begin{array}{ccc}
 {1 \over 2} & {1 \over 2} &~1  \\ \nonumber
 j_{b} & j_{a} & l_{a}, \end{array} \right\} \\ \nonumber
\end{eqnarray}
where the expansion coefficients for bra and ket vectors are included at the double expansion coefficient, $F_{{\alpha_p}a {\beta_n}b }^{1K}$, defined below Eq. (\ref{eq:gap}).
Finally, in order to compare our theoretical results to the experimental data, the GT($\mp$) strength functions and their running sums (total strengths) are calculated as
\begin{eqnarray} \label{eq:bgt}
&&B_{GT}^{-}(m)= \sum_{K=0,\pm 1} | < K^+ ,m | {\hat \beta}_{K}^- | ~QRPA > |^2, \\ \nonumber
&&B_{GT}^{+}(m)= \sum_{K=0,\pm 1} | < K^+,m | {\hat \beta}_{K}^+ | ~QRPA > |^2,\end{eqnarray}
\begin{eqnarray} \label{eq:str}
&&S_{GT}^{-}= \sum_{K=0,\pm 1} {\sum_{m}} | <  K^+,m | {\hat \beta}_{K}^- | ~QRPA > |^2,
\\ \nonumber
&&S_{GT}^{+}= \sum_{K=0,\pm 1} {\sum_{m}} | <  K^+,m | {\hat \beta}_{K}^+ | ~QRPA > |^2,
\end{eqnarray}
where $m$ is one of the $| K^+>$ intermediate states in odd-odd daughter nucleus. Numerical results and discussions are presented in section III.B with comparison to available experimental data.
\subsection{Ikeda Sum Rule}
In order to test our DQRPA model, numerical results for total GT$({\pm})$ strengths, $S_{GT}^{-}$ and $S_{GT}^{+}$, in Eq. (\ref{eq:str}) are investigated through the Ikeda sum-rule (ISR), which is known to be satisfied more or less
independently of the excited states constructed by any nuclear models,
\begin{equation}\label{eq:isr}
S_{GT}^{-} - S_{GT}^{+} = 3(N-Z).
\end{equation}
The ISR within the Deformed QRPA, denoted as ISR II, is given by
\begin{eqnarray}\label{eq:isr-d-II}
 && {(S_{GT}^{-} - S_{GT}^{+})}_{ISR~ II}  \nonumber \\
 && = \sum_{K=0,\pm 1} \sum_{m} [~ | <  K^+, m | {\hat \beta}_{K }^- | ~QRPA > |^2
- |<  K^+, m | {\hat \beta}_{K }^+ | ~QRPA > |^2 ] \nonumber \\
&&=  \sum_{K=0,\pm 1}\sum_{m} {\sum_{\alpha \rho_{\alpha} \beta \rho_{\beta}}} | < \alpha p \rho_{\alpha} |\tau^{+} \sigma_K
| \beta n \rho_{\beta}> |^2 ~ ( u^2_{\alpha p} v^2_{ \beta n}- v^2_{\alpha p} u^2_{ \beta n} )[ ( {X^m_{(\alpha p \beta n)K}})^2- ({Y^m_{(\alpha p \beta n)K}})^2 ]~. \nonumber \\
\end{eqnarray}
If we use the closure relation for the excited states, the ISR which we denote it as ISR I is shown to be easily calculated as follows
\begin{eqnarray}\label{eq:isr-d-I}
&& {(S_{GT}^{-} - S_{GT}^{+})}_{ISR~ I} \nonumber \\
&&= \sum_{K=0,\pm 1} {\sum_{\alpha \rho_{\alpha} \beta \rho_{\beta}}}
|< \alpha p \rho_{p} | \tau^{+} \sigma_{K}| \beta n \rho_{n} >|^2 (v_{n}^{2}-v_{p}^{2}) ~.
\end{eqnarray}
Since we use the expansion of a deformed wave function in terms of the single particle basis,
the above sum rule might be a bit broken at $|< \alpha p \rho_{p} | \tau^{+} \sigma_{K}| \beta n \rho_{n} >|^2 $ in Eqs. (\ref{eq:isr-d-I}) and (\ref{eq:isr-d-II}). But it could be a useful tool to assess our nuclear model and numerical calculations. In particular, ISR II could be a meaningful verification of the DQRPA calculation while ISR I might be a good test for the DBCS calculation. Our results of ISR I and II are tabulated in Table \ref{tab:parameters} with detailed discussions at section III.A. Of course, the ISR in the spherical QRPA is easily shown to satisfy the ISR as follows
\begin{eqnarray}\label{eq:isr-s}
S_{GT}^{-} - S_{GT}^{+}
&&=\sum_{a b}
|< a p | \tau^{+} \sigma | b n >|^2 (v_{n}^{2}-v_{p}^{2}) \\\nonumber
&&= \sum_{a b}(2j_{a}+1)(2j_{b}+1)
\delta_{n_{a} n_{b}} \delta_{l_{a} l_{b}}
\left \{ \begin{array}{ccc}
 {1 \over 2} & {1 \over 2} &~1  \\ \nonumber
 j_{b} & j_{a} & l_{a} \end{array} \right \}^2
(v_{n}^{2}-v_{p}^{2})  \\ \nonumber
&& =3 \sum_{b} (2j_{b}+1) v_{n}^{2} - 3 \sum_{a}(2j_{a}+1) v_{p}^{2} = 3(N-Z).
\end{eqnarray}
\section{Numerical Results and Discussions}
We calculated the Gamow-Teller (GT) strength distributions, B$^{\pm}_{GT}$ in Eq. (\ref{eq:bgt}) and their running sums S$^{\pm}_{GT}$ in Eq. (\ref{eq:str}), for $^{76}$Ge, $^{76,82}$Se, $^{90}$Zr, and $^{92}$Zr within the DQRPA. Those nuclei are selected to represent medium-heavy nuclei which have experimental data of the GT strength distributions.
Before showing the numerical results, we discuss how to fix physical parameters associated to this work.

\subsection{Deformation, Pairing strength, Particle-particle and Particle-hole strength parameters, and Ikeda Sum Rules}

\subsubsection{Deformation parameter}
The single particle state energies adopted from the deformed Woods-Saxon potential naturally depend on the deformation parameter $\beta_2$. The deformation of nuclei may be conjectured to be closely associated with macroscopic phenomena, for example, the core polarization, the high spin states and so on. Microscopically it may result from the tensor force in the N-N interaction, which is known to account for the shell evolution according to the recent systematic shell model calculations \cite{Otsuka05,Otsuka10}.  For example, T = 0, J = 1 pairing, which is associated with the $^{3}S_0$ tensor force, may lead to the deformation contrary to the spherical T = 1, J = 0 pairing. Therefore, the deformation parameters adopted in this work may include implicitly and effectively such effects, because the single particle states from the deformed Woods-Saxon potential show also a strong dependence on the $\beta_2$ \cite{Nilsson}. For realistic example,  in our previous paper \cite{Ha13_1}, the Woods-Saxon (Coulomb corrected) potential used in this work was shown to explain the shell evolution of Mg isotopes.

The deformation parameter $\beta_2$ helped us to conjecture the nuclear shape through the intrinsic quadrupole moment, $Q = \sqrt{16 \pi /5}(3/4\pi)A R_0^2 \beta_2$,
where $R_0 =1.2 A^{1/3}$fm for the sharp-cut radius $R_0$ \cite{Hagino, beta2}.
In the experimental side, the deformation parameter $\beta_2$
can be extracted from the E2 transition probability, $Q = \sqrt{ 16 \pi B(E2)/5e^2 }$, which we denote it as $\beta_2^{E2}$ \cite{Raman}.
For a reference, theoretical values of the $\beta_2$ by the relativistic mean field (RMF) theory \cite{Lala99}, denoted as $\beta_2^{RMF}$, are indicated together in Table \ref{tab:deformation}. But signs of the $\beta_2$ are sometimes still uncertain. Therefore we exploited various values of the deformation parameter, $0.1 \le |\beta_2| \le 0.3$, as default values. In this work, the coexistence of prolate and oblate shapes and the $\beta_4$ deformation are not taken into account.
\subsubsection{Pairing strength parameters}
For the pairing interaction, the strength parameters $g_{\textrm{pair}}^{n}$ and $g_{\textrm{pair}}^{p}$ in Eq. (\ref{eq:gap}),
which are introduced to renormalize the finite Hilbert model particle space, are adjusted to reproduce empirical
pairing gaps through the symmetric five term formula in Eqs. (\ref{eq:gap_p}) and (\ref{eq:gap_n}) \cite{Ch93}. Results of strength parameters $g_{\textrm{pair}}^{n}$ and $g_{\textrm{pair}}^{p}$ for a given $\beta_2$ value are recapitulated with theoretical and empirical pairing gaps in Table \ref{tab:parameters}.

\begin{figure}
\includegraphics[width=0.8\linewidth]{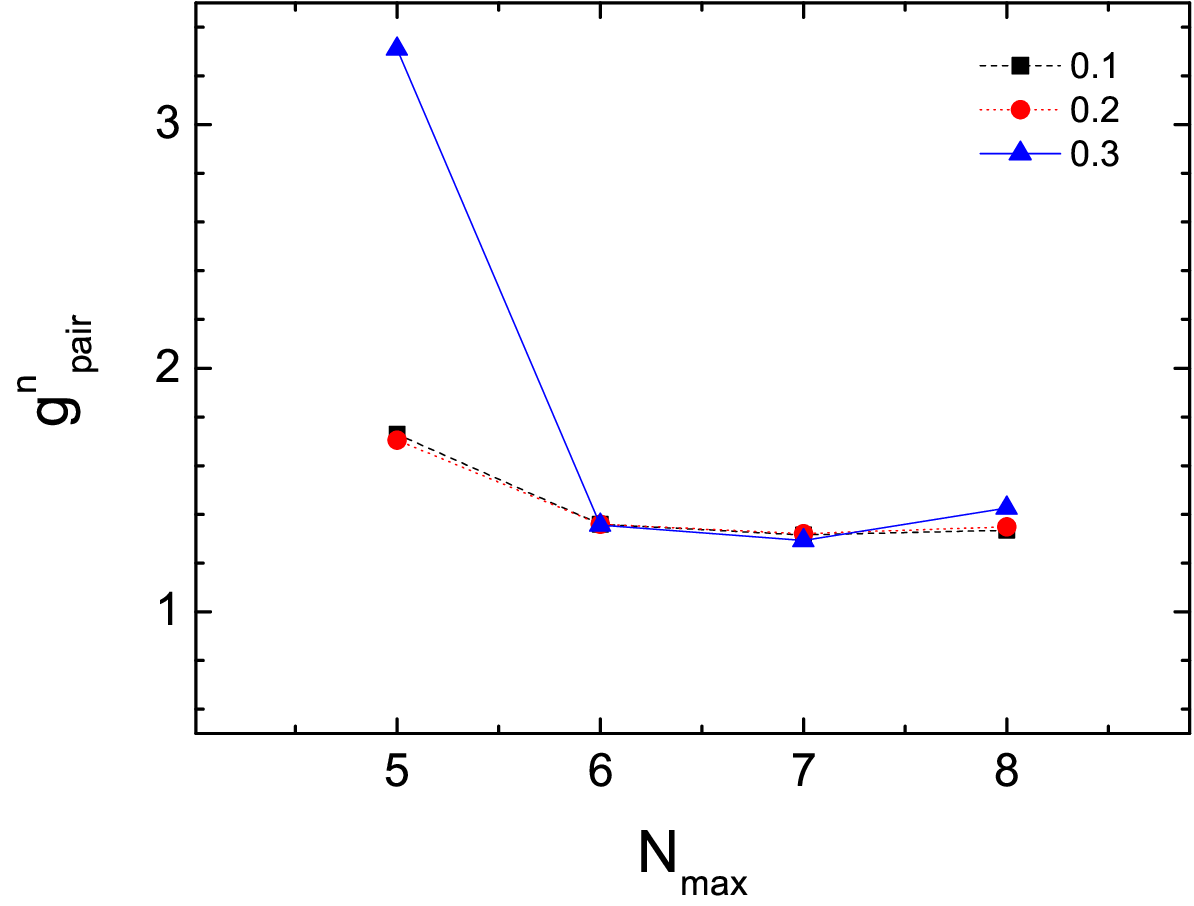}
\caption{(Color online) Dependence of neutron pairing strength $g_{pair}^{n}$ in Eq. (\ref{eq:gap}) on particle model space $N_{max}^{sph}$ in $G-$matrix.
Black dashed, red dotted, and blue solid points are results for $\beta_2$ = 0.1, 0.2, and 0.3, respectively.}
\label{fig1}
\end{figure}

In Fig. \ref{fig1}, we briefly discuss dependence of the strength parameter $g_{\textrm{pair}}^{n(p)}$ on particle model space. If we use $N_{max}^{sph} = 5 \hbar\omega $ for the particle model space in $G-$matrix, the strength parameter $g_{\textrm{pair}}^{n}$ is abnormally deviated from 1. For instance, it may go beyond 2 for $\beta_2 = 0.3$ (blue solid line) for $^{76}$Ge as shown in Fig. \ref{fig1}. It means that the particle model space $N_{max}^{sph} = 5\hbar\omega $ is not enough to reproduce the empirical pairing gaps $\Delta_{em}^{n(p)}$. Therefore, in this calculation, we use $N_{max}^{sph}$=$10 \hbar \omega$ in $G-$matrix, which enables us to obtain $g_{\textrm{pair}}^{n} ( g_{\textrm{pair}}^{p}$) values around 1.1 $\sim$ 1.6
as shown in Table \ref{tab:parameters}. Generally, $g_{\textrm{pair}}^n$ is larger than $g_{\textrm{pair}}^p$. It means that, for stable nuclei which have more neutrons, neutron model spaces are not relatively enough compared to proton model spaces if we use the same $N_{max}$ for both particle spaces.

\subsubsection{Particle-particle and particle-hole strength parameters}
\begin{figure}
\includegraphics[width=0.8\linewidth]{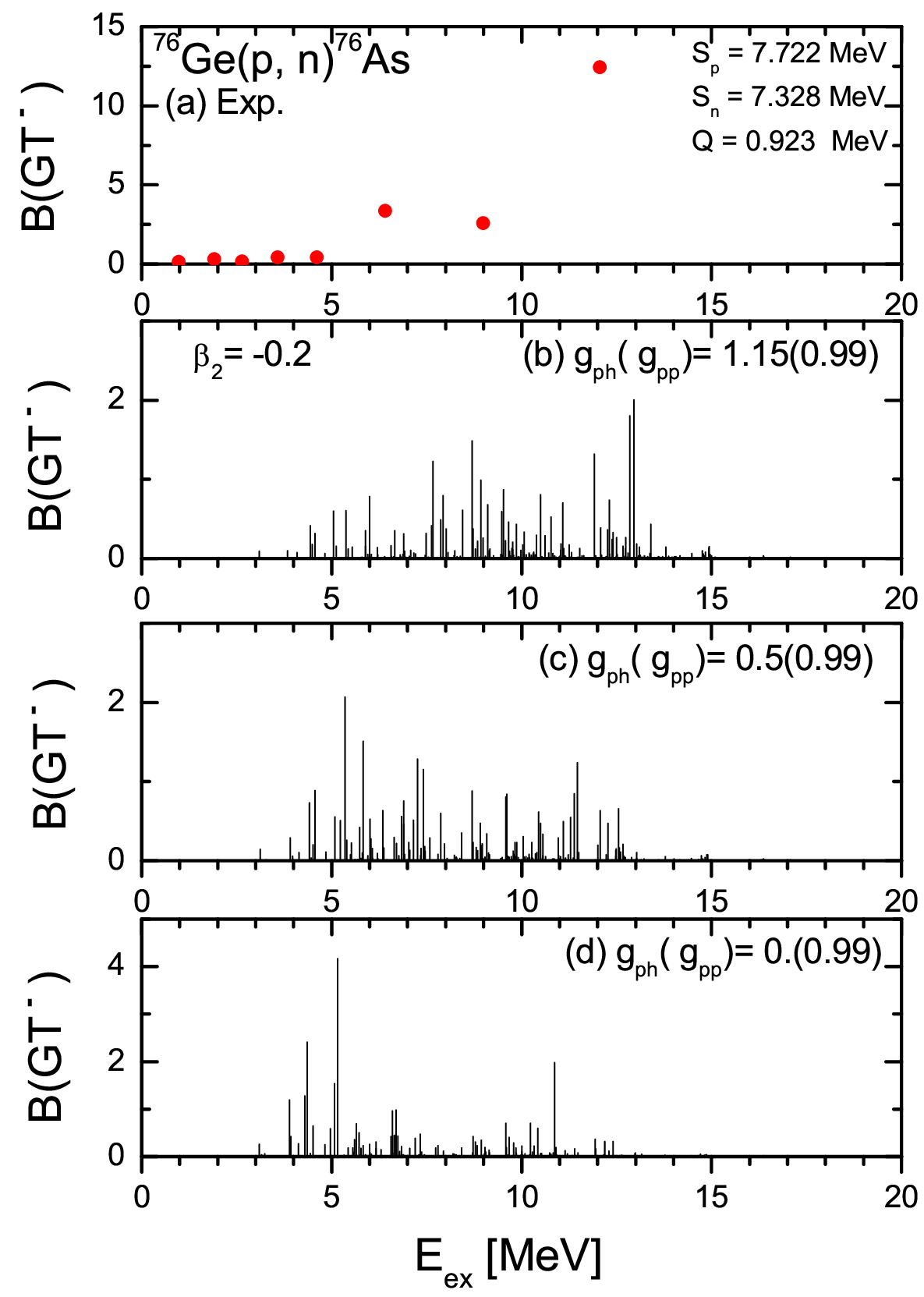}
\caption{(Color online) GT strength distributions in $^{76}$Ge for different particle-hole interaction strength parameter $g_{ph}$ in Eq. (\ref{eq:rpa_a}) - (\ref{eq:rpa_b}) as a function of the excitation energy $E_{ex}$ w.r.t a parent nucleus. Experimental data denoted as filled (red) points in the uppermost panel are deduced from the $^{76}$Ge(p,n) reaction at 134.4 MeV \cite{Madey89}.
The particle-particle interaction strength $g_{pp}$ is fixed as 0.99 for (b) - (d).}
\label{fig2}
\end{figure}

In the DQRPA stage, we took the
particle-hole and particle-particle strength parameters, $g_{ph}$ and $g_{pp}$, in Eq. (\ref{eq:mat_A}) -- (\ref{eq:rpa_b}) as 1.15 and 0.99 for all nuclei. Actually $g_{ph}$ might be determined from the Gamow Teller Giant Resonance (GTGR), while fine tuning of $g_{pp}$ is usually performed for the double beta decay \cite{Ch93}. In order to grasp the $g_{ph}$ dependence, in Fig. \ref{fig2}, the GT strength distributions for $^{76}$Ge for a temporally fixed $\beta_2 = - 0.2$ are shown for different $g_{ph} =$ 0, 0.5, and 1.15, with a fixed $g_{pp} = 0.99$. One may notice that positions of the GTGR are sensitive to the $g_{ph}$ value. $g_{ph}$=1.15 ((b) panel) reproduce the position of the GTGR energy, which is also consistent with other calculations \cite{saleh}.

\begin{figure}
\includegraphics[width=0.8\linewidth]{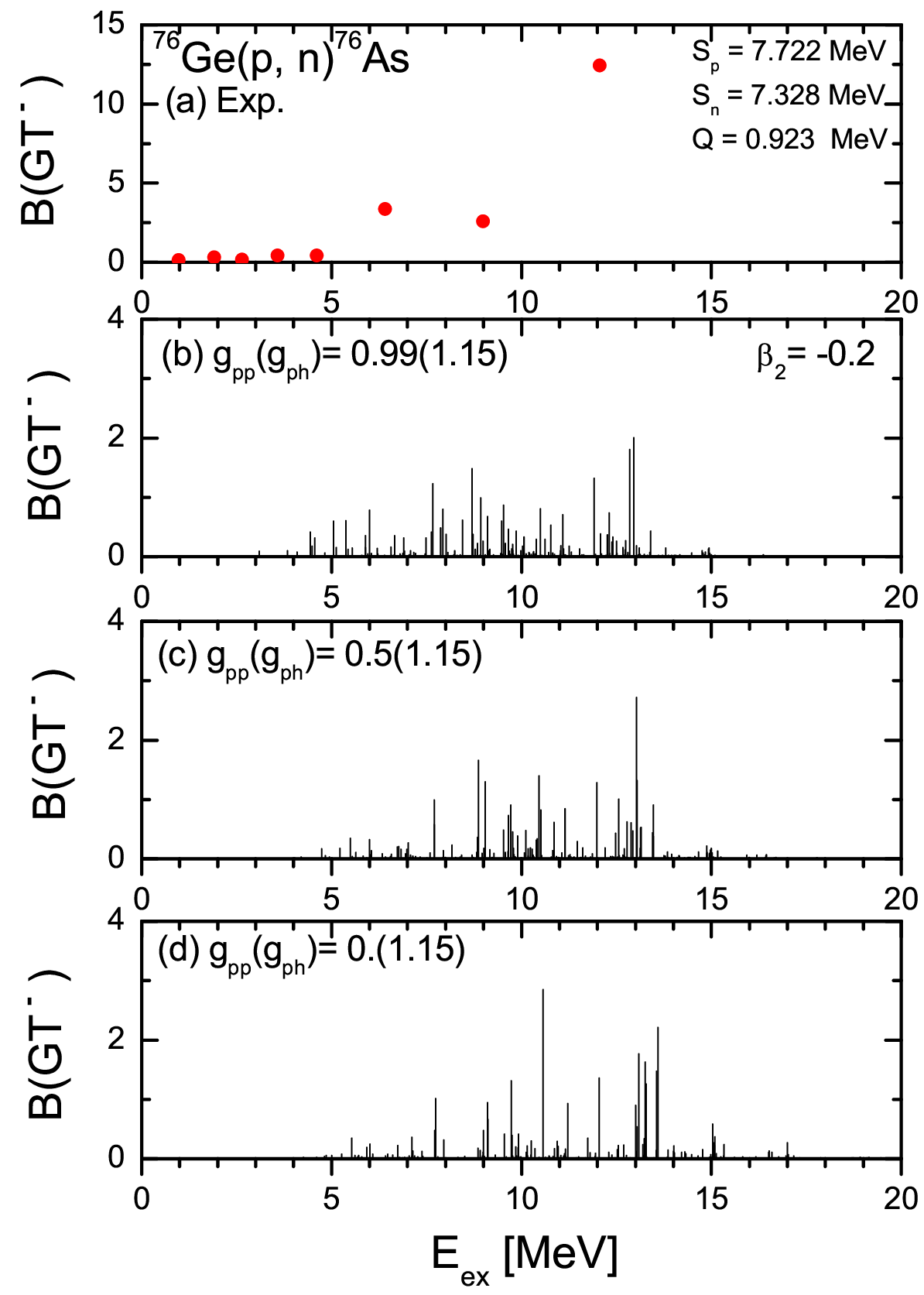}
\caption{(Color online) GT strength distributions in $^{76}$Ge for different particle-particle interaction strength parameter $g_{pp}$ in Eq. (\ref{eq:rpa_a}) - (\ref{eq:rpa_b}) as a function of the excitation energy $E_{ex}$.
The particle-hole interaction strength $g_{ph}$ is fixed as 1.15 for (b) - (d).}
\label{fig3}
\end{figure}

Fig. \ref{fig3} illustrates the evolution of the GT strength distributions with respect to the increase of $g_{pp}$= 0, 0.5, and 0.99,
with a fixed $g_{ph} = 1.15$. One can see that all GT peaks get shifted only to a small amount of energies as $g_{pp}$ approaches to 0.99.

Since the nuclei considered here are expected to have large energy gaps between proton and neutron spaces,
we considered only the $nn$ and $pp$ pairing correlation although the formalism is presented generally for further applications to neutron-deficient (proton-rich) nuclei.
For example, in the neutron-rich nuclei of importance in the r-process, the {\it np} pairing may not contribute so much.
But for the p-process, the $np$ pairing could become important because of the adjacent
energy gaps of protons and neutrons. Calculations for the neutron-deficient nuclei in the p-process are in progress by the explicit inclusion of the {\it np} pairing correlations.

\subsubsection{Ikeda Sum Rule}
Last column in Table \ref{tab:parameters} is results of the ISR I and II, Eqs. (\ref{eq:isr-d-I}) and (\ref{eq:isr-d-II}). All of ISR results are well satisfied in our nuclear model, and nearly
independent of the deformation parameter, $\beta_2$, as required in the sum rule. If the single particle states are used up to $N_{max}^{def} = 6 \hbar \omega$ for all nuclei, ISR could be satisfied beyond 99$\%$. Another point to be noticed is that both ISR I and II results are exactly coincident with each other less than 1\%.

\begin{figure}
\includegraphics[width=0.8\linewidth]{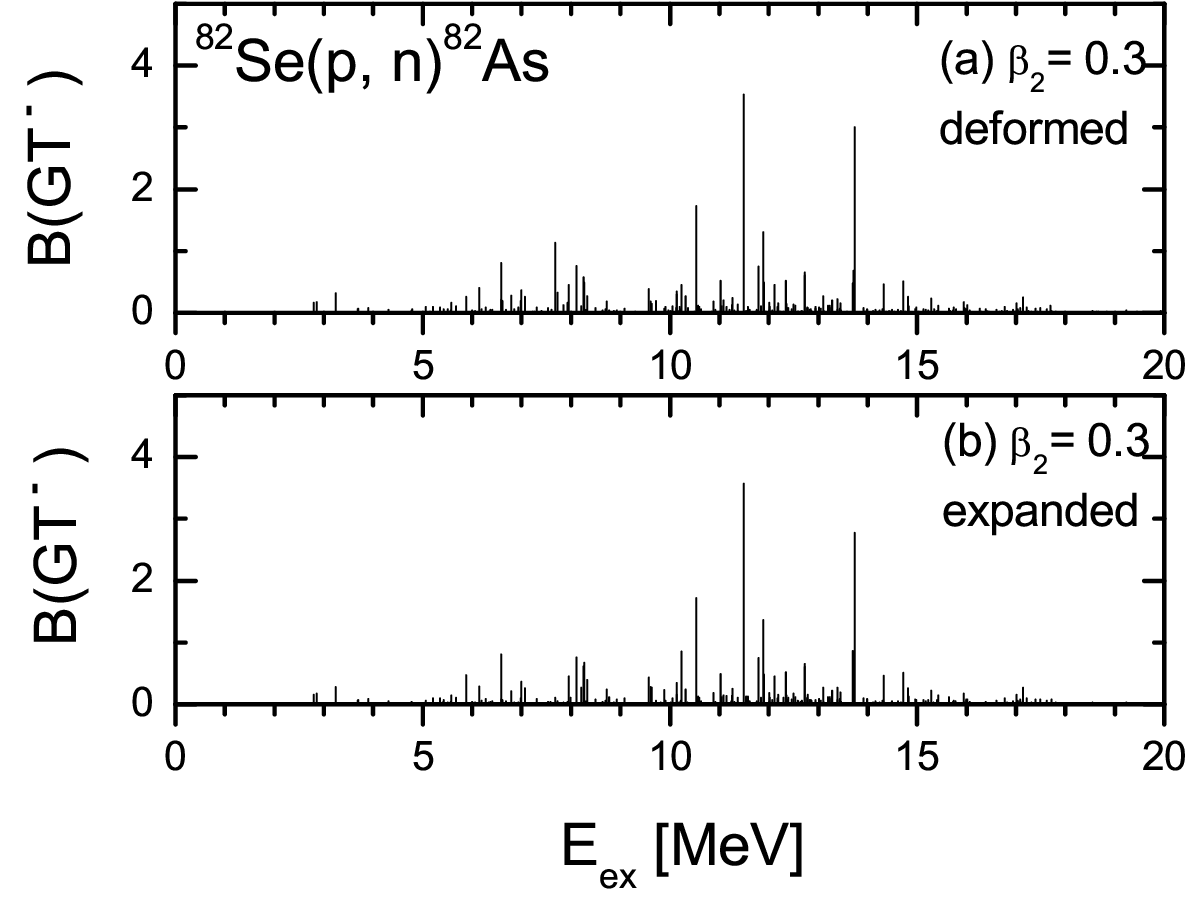}
\caption{(Color online) GT strength distributions for $^{82}$Se at $\beta_2 = 0.3.$ They are calculated in deformed basis by Eq. (\ref{eq:tram_d1})$\sim$(\ref{eq:tram_d3}) (a) and in the spherical basis by Eq. (\ref{babo}) (b).}
\label{fig4}
\end{figure}

ISR I and II in Table \ref{tab:parameters} are calculated by two different ways. First, we calculate in the spherical basis with the expansion method in Eq. (\ref{babo}). Second, we do them in the deformed basis using Eq. (\ref{eq:tram_d1}) -- (\ref{eq:tram_d3}).
 Both ISR results are confirmed to be equal to each other within a few percentage deviation. Results of the GT strength distributions also show identical results irrespective of the two different approaches adopted, as shown in Fig. \ref{fig4}, where GT strength distribution results for $^{82}$Se at $\beta_2 =0.3$ are shown for the case calculated by the deformed basis (a) and the case by the expansion method (b).

\subsection{Comparison of theoretical GT strength distributions with available experimental data}

In the following, we discuss the GT ($\mp$) strength distribution function by showing our numerical results for $^{76}$Ge, $^{76,82}$Se, $^{90,92}$Zr, which are presented with respect to the excitation energy, $E_{ex}$, of parent nuclei. Since most of the GT strength data are reported by the $E_{ex}$ of daughter nuclei, the experimental data in the following figures are shown by adding the empirical Q values ($Q_{\beta^-}, Q_{EC}$) from the measured data $E_{ex}$. In particular, we focus on roles of the deformation parameter $\beta_2$ in the GT ($\mp$) strength distribution.

\subsubsection{$^{76}$Ge}
\begin{figure}
\includegraphics[width=0.8\linewidth]{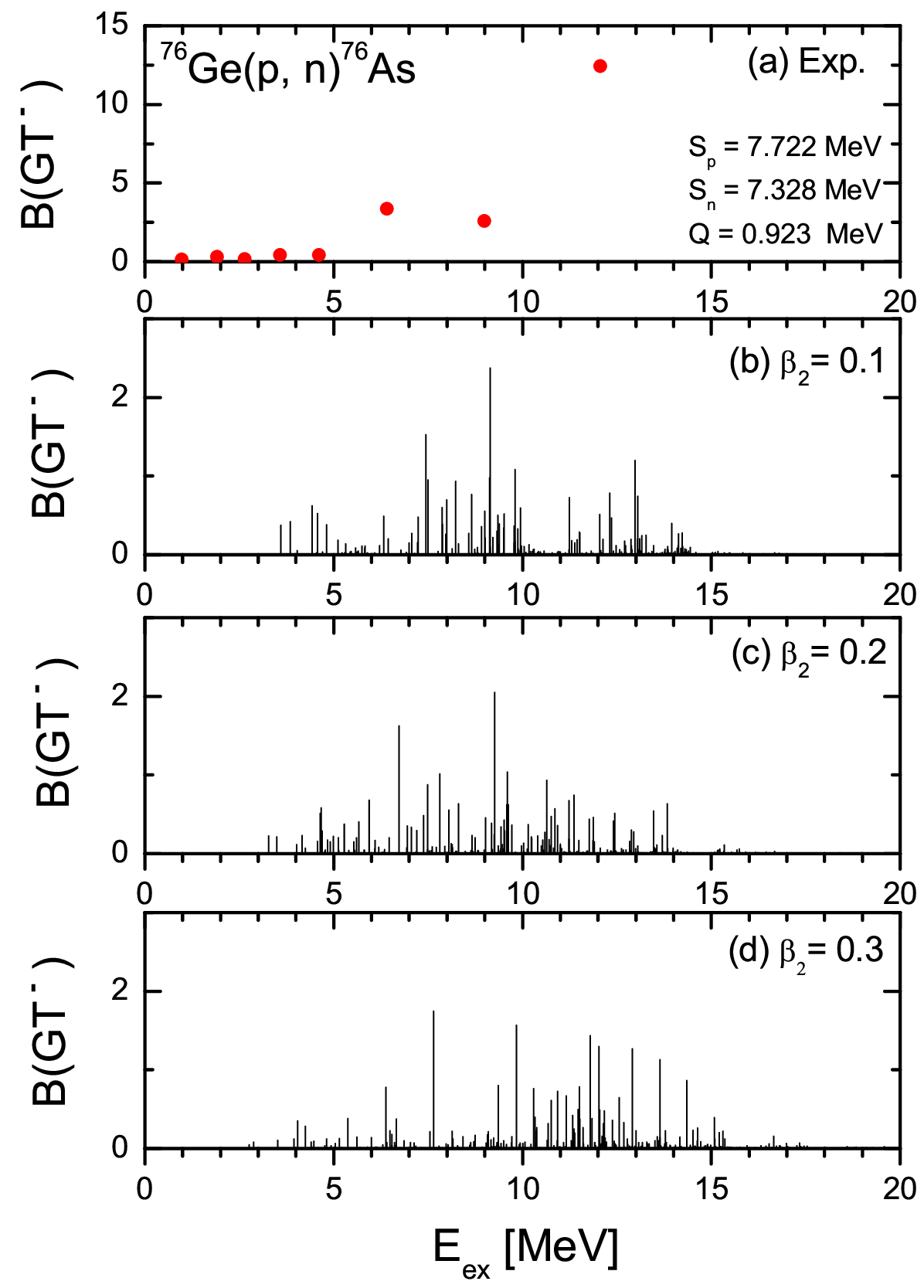}
\caption{(Color online) GT strength distributions B(GT$^{-}$)
on $ ^{76}$Ge as a function of the excitation energy $E_{ex}$ w.r.t. the ground state of $^{76}$Ge.
Experimental data denoted as filled (red) points in the uppermost panel are deduced from the $^{76}$Ge(p,n)
reaction at 134.4 MeV \cite{Madey89}. In each panel, we indicate each $\beta_2$ value (0.1 $\sim$ 0.3) for prolate shapes.
$\beta_{2}^{RMF}$= 0.157 and $\beta_{2}^{E2}$= 0.2623 (39) for $^{76}$Ge \cite{Lala99, Raman}.
}
\label{fig5}
\end{figure}

\begin{figure}
\includegraphics[width=0.8\linewidth]{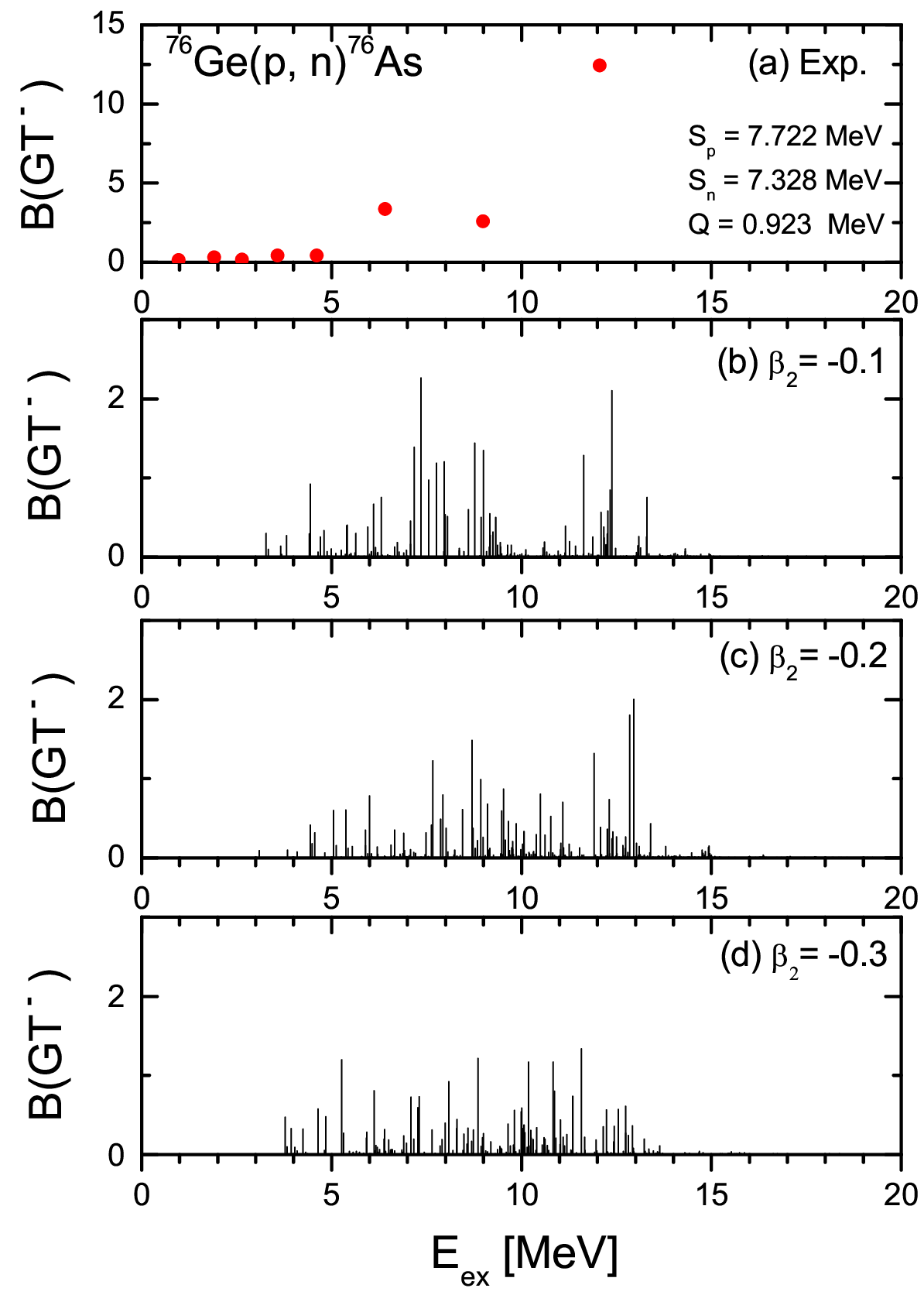}
\caption{(Color online) The same as in Fig. \ref{fig5}, but for oblate shapes, $\beta_2 = -0.1 \sim -0.3$.}
\label{fig6}
\end{figure}

In Figs. \ref{fig5} and \ref{fig6}, we show the GT(--) strength of $^{76}$Ge as a
function of the excitation energy $E_{ex}$ w.r.t. the ground state of $^{76}$Ge, whose Q value and proton (neutron) separation energy are 0.9233 MeV and 7.722 (7.328) MeV, respectively. The uppermost panels are the experimental data from the $^{76}$Ge(p,n)$^{76}$As reaction at 134.4 MeV \cite{Madey89},
which show a strong GT state peak around 12 MeV. Fig. \ref{fig5} represents the GT strength distributions with prolate shapes, $\beta_2 = 0.1 \sim 0.3$,
and Fig. \ref{fig6} is for oblate shapes, $\beta_2 = -0.1 \sim -0.3$.
The GT strength distributions are widely scattered owing to the deformation. If we look in detail the GT strength distribution in Figs. \ref{fig5} and \ref{fig6}, the GTGR appears around 12 MeV for $\beta_2 = 0.3, -0.1, -0.2$. But results of the oblate types in Fig. \ref{fig6} indicate relatively strong strength around 9 MeV contrary to the experimental data.

\begin{figure}
\includegraphics[width=0.8\linewidth]{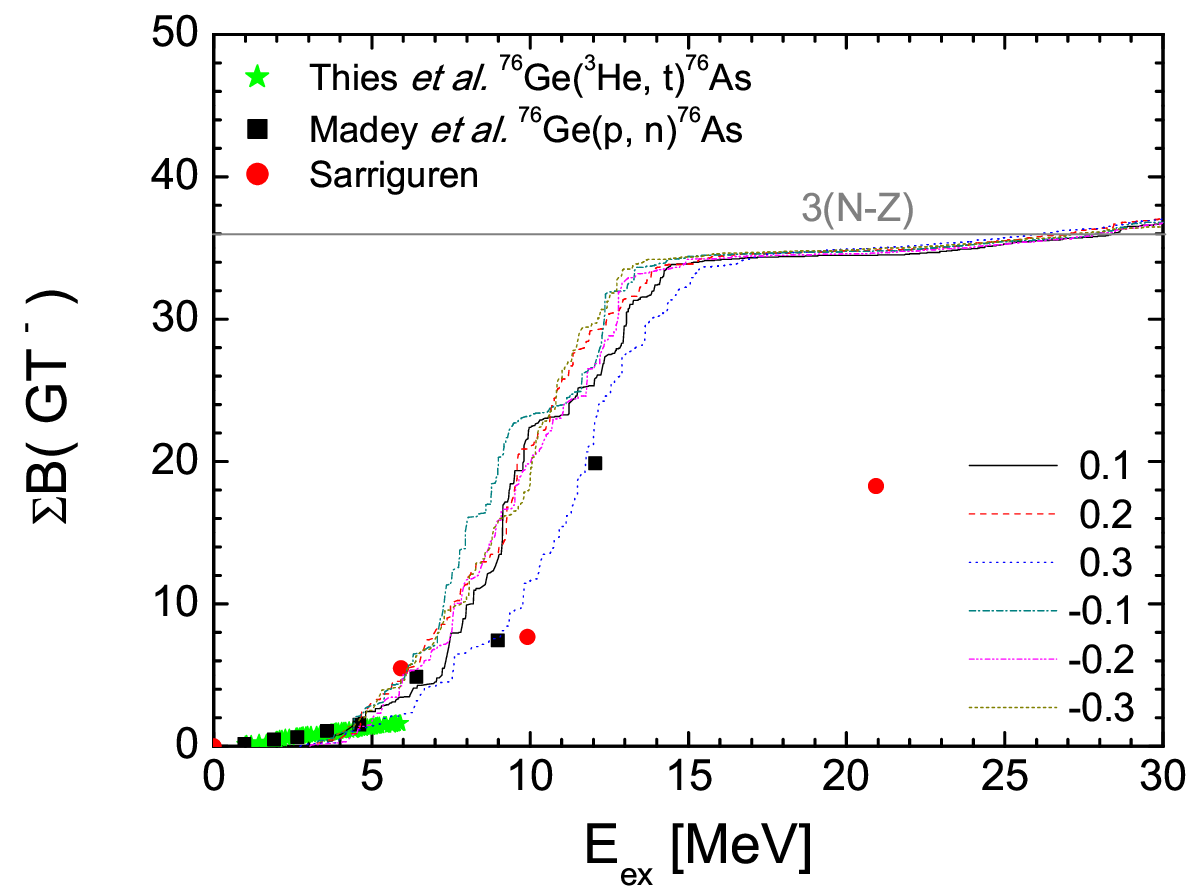}
\caption{(Color online)  Running sums for the GT (--) strength distributions (a) - (d) in Figs. \ref{fig5} and \ref{fig6}. Results by Thies {\it et al.} and Madey {\it et al.} are experimental data deduced from Refs. \cite{Thies} and \cite{Madey89}. Red dotted points are theoretical results at Table II in Ref. \cite{sarriguren_d}.}
\label{fig7}
\end{figure}

Therefore, in Fig. \ref{fig7}, in order to understand more clearly the $\beta_2$ dependence, we present the running sum of GT strength, Eq. (\ref{eq:str}), on $^{76}$Ge calculated up to 50 MeV for different $\beta_2$ values, $\pm 0.1\sim \pm 0.3$,  as a function of the excitation energy $E_{ex}$.
In particular, the results for $\beta_2$ = 0.3 represented by blue dotted line nicely reproduce the experimental data. Therefore $\beta_2 = 0.3$ seems to be most appropriate for $^{76}$Ge to explain both strength distribution functions and their running sums of the GT strength. Our $\beta_2$ value is more or less consistent with the prolate shape value extracted from B(E2), whose $\beta_2$ value is 0.2623 (39) \cite{Raman}. For a reference, the $\beta_2$ value from the RMF is 0.157 \cite{Lala99}.

Our results for the running sum up to 30 MeV are 37.49(37.22), 37.60(37.04), and 37.65(36.77) for $\beta_2$= 0.1(-0.1), 0.2(-0.2), and 0.3(-0.3) respectively. They satisfy about 98\% of the ISR, 3(N-Z) = 36, if we take the running sum of the B(GT+) in Fig. \ref{fig10}. If we look at the experimental running sum data in Fig. \ref{fig7}, the running sum up to 6 MeV and 12 MeV are 1.6 \cite{Thies} and 19.9 \cite{Madey89}, respectively. It means that the ISR is recovered about 55 \% in the experiment as argued in Ref. \cite{Madey89}, if the $\beta^+$ strength is assumed to be zero.
By using the universal quenching factor 0.79 for the axial coupling constant $g_A$, one may compromise theoretical calculations to the underestimated running sum data. But we conjecture that there may be a possibility of the high-lying GT state above 12.0 MeV as shown in Fig. \ref{fig5} (d), which can compensate significantly the underestimated experimental data. For a reference, the running sum calculated by the DQRPA with the Skyrme force by Sarriguren is 18.3 up to 20 MeV  \cite{sarriguren_d} indicated as red points in Fig. \ref{fig7}.

\subsubsection{$^{76}$Se}
\begin{figure}
\includegraphics[width=0.8\linewidth]{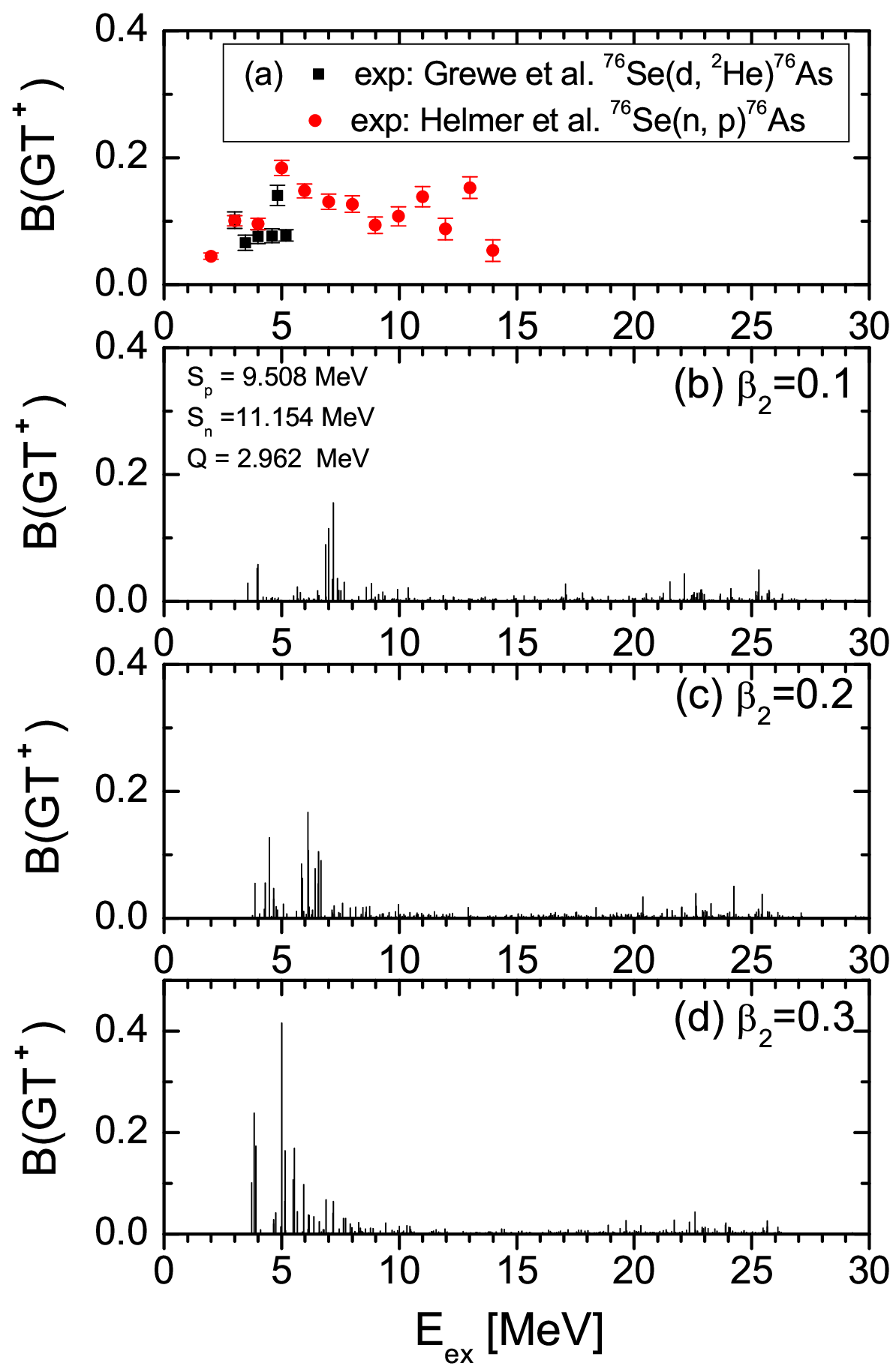}
\caption{(Color online) GT strength distributions B(GT$^{+}$)
on $ ^{76}$Se as a function of the excitation energy $E_{ex}$ w.r.t. the ground state of $^{76}$Se.
Experimental data denoted as red points and black squares in the uppermost panel are deduced from the $^{76}$Se(n,p)
reaction at 134.4 MeV \cite{Helmer} and $^{76}$Se(d,$^{2}$He) \cite{Grewe}. In each panel, we indicate each $\beta_2$ value (0.1 $\sim$ 0.3) for prolate shapes.
$\beta_{2}^{RMF}$= -0.244 and $\beta_{2}^{E2}$= 0.309 (37) for $^{76}$Se \cite{Lala99, Raman}.
}
\label{fig8}
\end{figure}

\begin{figure}
\includegraphics[width=0.8\linewidth]{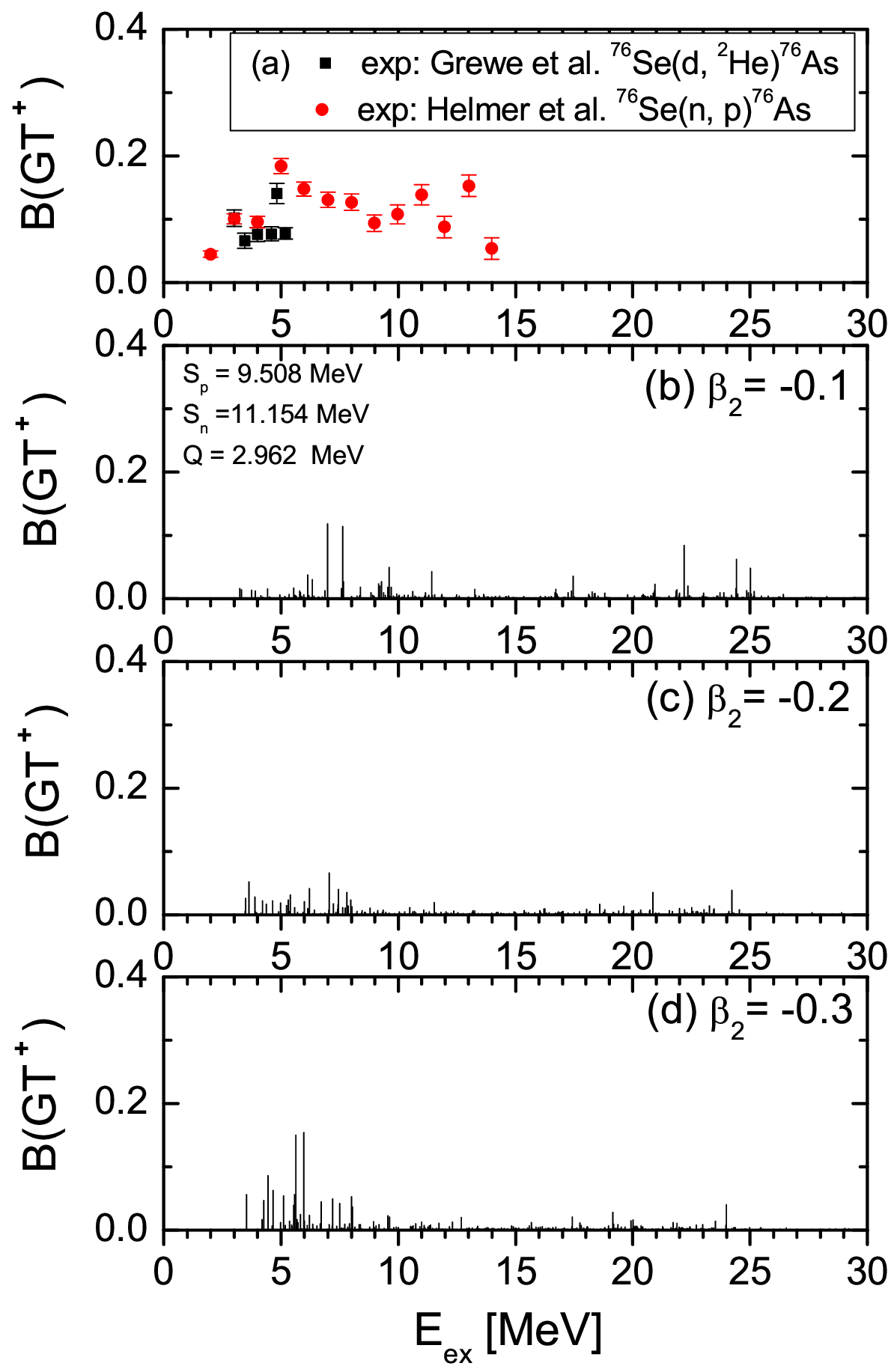}
\caption{(Color online) The same as in Fig.\ref{fig8}, but for oblate shapes, $\beta_2 = -0.1 \sim -0.3$.}
\label{fig9}
\end{figure}

\begin{figure}
\includegraphics[width=0.8\linewidth]{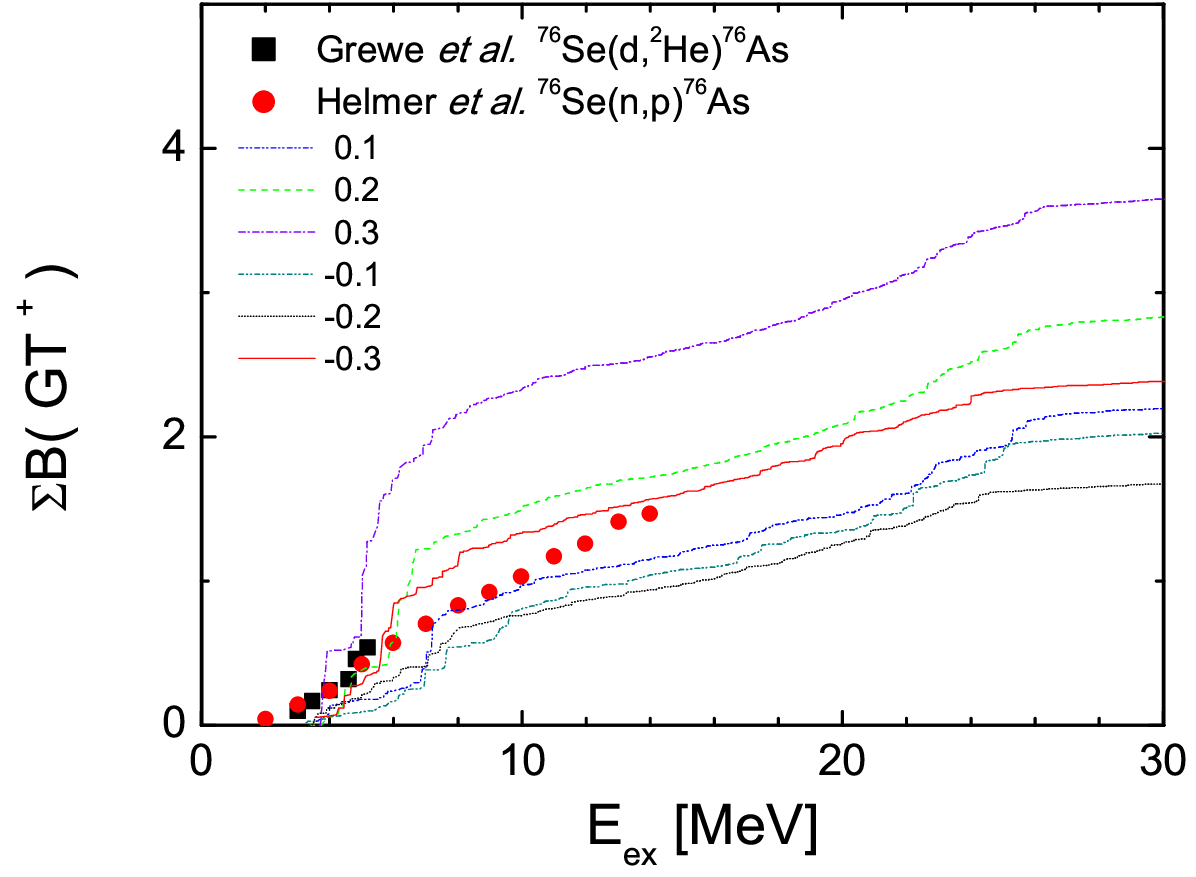}
\caption{(Color online) Running sums for the GT(+) strength distributions (a) - (d) in Figs. \ref{fig8} and \ref{fig9}.
Results by Grewe {\it et al.} and Helmer {\it et al.} are experimental data deduced from Refs. \cite{Grewe} and \cite{Helmer}.}
\label{fig10}
\end{figure}

In a similar way, the GT(+) strength distributions, B(GT$+$), for $^{76}$Se are presented in Figs. \ref{fig8} and \ref{fig9} as a
function of the $E_{ex}$ w.r.t. the ground state of $^{76}$Se (whose Q value is 2.962 MeV) for $\beta_2 = \pm 0.1 \sim \pm 0.3$.
The uppermost panels of Figs. \ref{fig8} and \ref{fig9} are the experimental data from the $^{76}$Se(n,p)$^{76}$As \cite{Helmer} and $^{76}$Se(d,$^{2}$He)$^{76}$As \cite{Grewe} with an improved energy resolution of 120 keV.
Lower panels show the deformed QRPA results for different values of the quadrupole deformation.
One may notice that the main GT peak gets shifted to lower energy region as the $|\beta_2|$ value increases for both prolate and oblate shapes. Since we could not decide which shape is the better of the two shapes from the GT(+) distributions, we compare their running sums up to 30 MeV in Fig. \ref{fig10} for $\beta_2 = \pm 0.1\sim \pm 0.3$ as a function of the $E_{ex}$.
The result for $\beta_2 = - 0.3$ (red solid line) seems to be much better than other cases, if compared to the experimental data. Our $\beta_2$ value is consistent with the oblate shape extracted from the RMF theory, whose $\beta_2$ value is -- 0.244. For a reference, the $\beta_2^{E2} = 0.309(37)$.

\subsubsection{$^{82}$Se}
\begin{figure}
\includegraphics[width=0.7\linewidth]{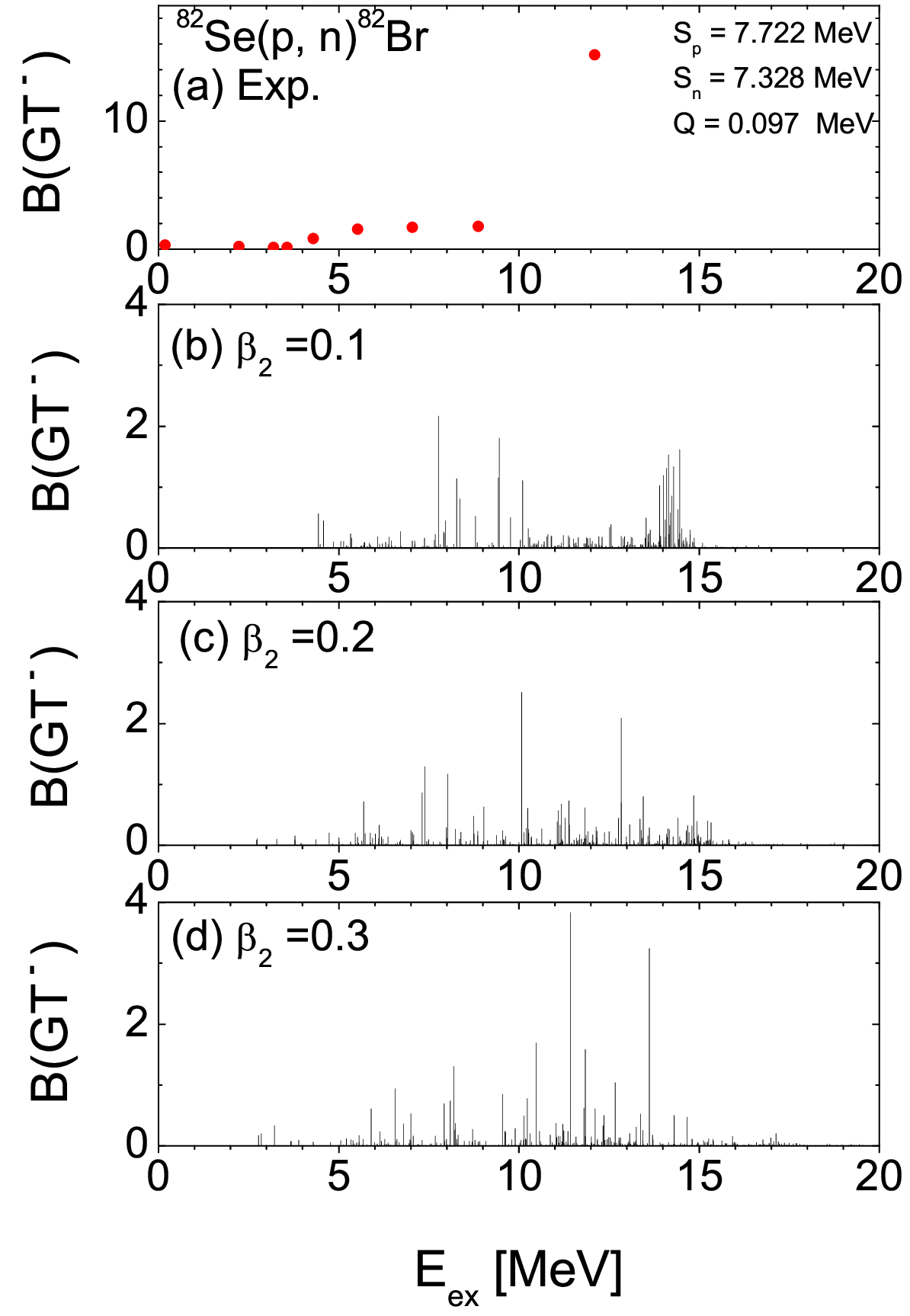}
\caption{(Color online) Gamow-Teller strength distributions B(GT$^{-}$) on $ ^{82}$Se as a function of the excitation
energy $E_{ex}$ w.r.t. the ground state of $^{82}$Se. Experimental data denoted as filled (red) points in the uppermost
panels are deduced from the $^{82}$Se(p,n) reaction at 134.4 MeV \cite{Madey89}. In each panel, we indicate $\beta_2$.
$\beta_{2}^{RMF}$= 0.133 and $\beta_{2}^{E2}$= 0.193 (27) for $^{82}$Se \cite{Lala99, Raman}.
}
\label{fig11}
\end{figure}

\begin{figure}
\includegraphics[width=0.7\linewidth]{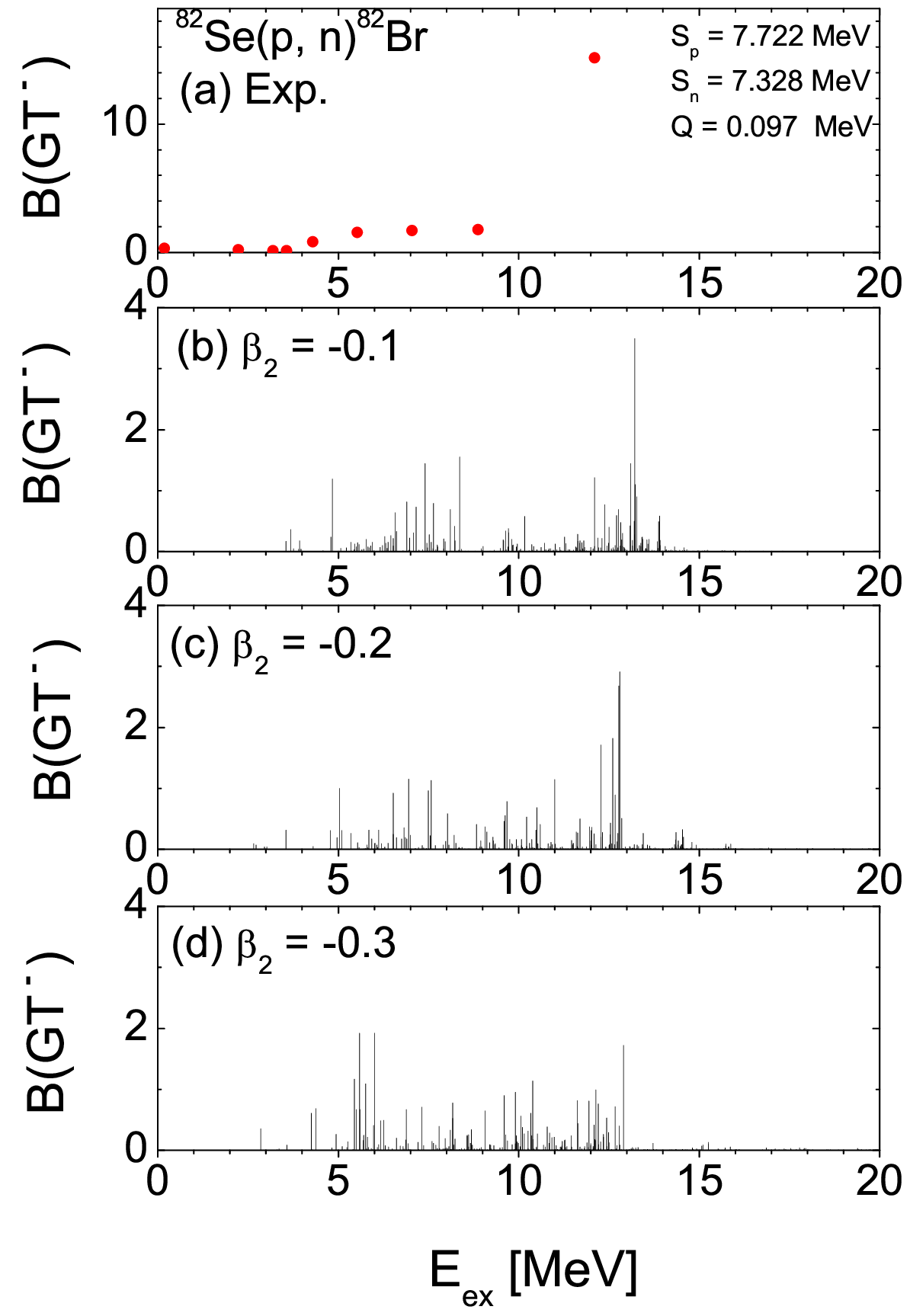}
\caption{(Color online) The same as in Fig.\ref{fig11}, but for oblate shapes, $\beta_2 = -0.1 \sim -0.3$.}
\label{fig12}
\end{figure}

\begin{figure}
\includegraphics[width=0.8\linewidth]{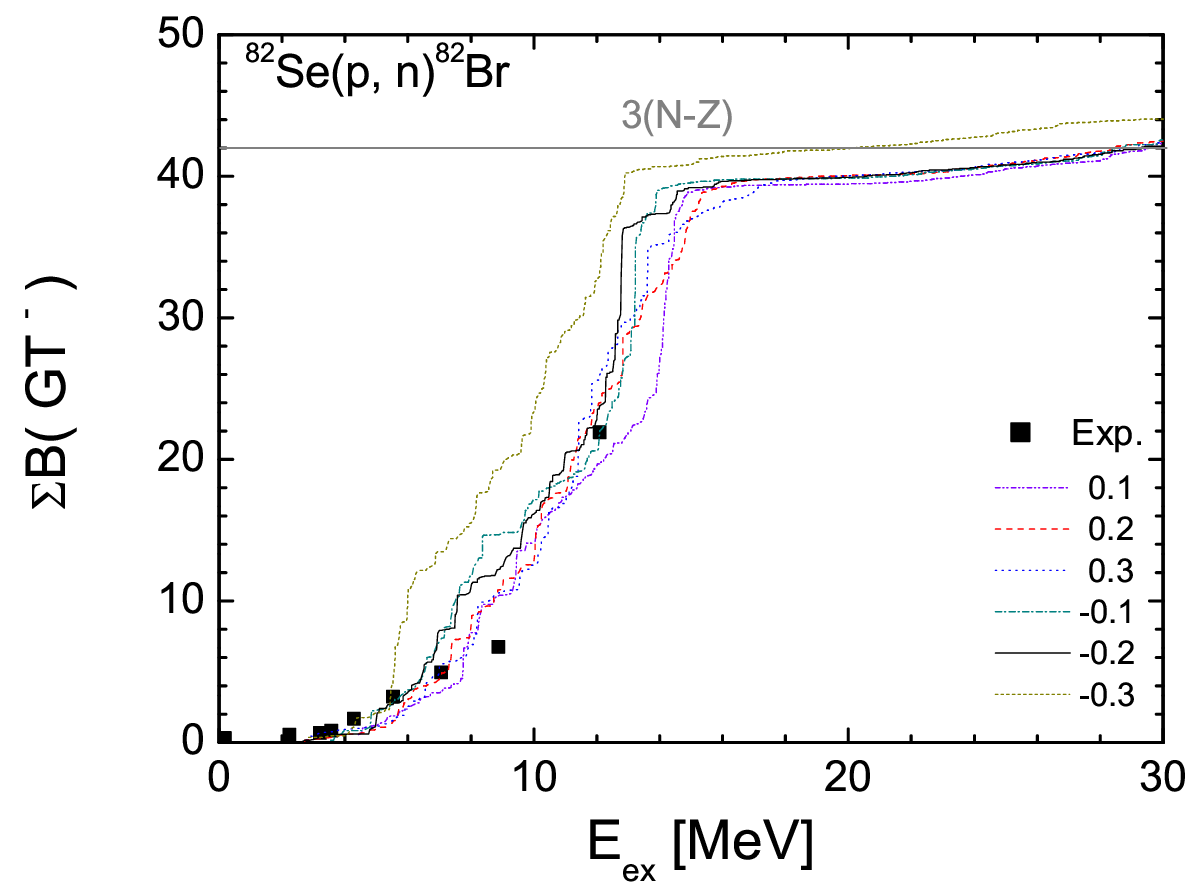}
\caption{(Color online) Running sums for the GT strength distributions (a) - (e) in Figs. \ref{fig11} and \ref{fig12}.
Experimental data are deduced from Ref. \cite{Madey89}.}
\label{fig13}
\end{figure}

The GT(--) strength distributions, B(GT$-$), on $^{82}$Se are presented, in Figs. \ref{fig11} and \ref{fig12}, as a
function of the $E_{ex}$ w.r.t. the ground state of $^{82}$Se (whose Q value is 0.092 MeV) for $\beta_2 = \pm 0.1 \sim \pm 0.3$.
The uppermost panels of Figs. \ref{fig11} and \ref{fig12} are the experimental data
from the $^{82}$Se(p,n)$^{82}$Br reaction at 134.4 MeV \cite{Madey89}.
Particularly, strengths of the GT excited state around 12 MeV are reproduced by the $\beta_2$ = 0.3 at (d) panel in Fig. \ref{fig11}, which was also confirmed at the running sum of the GT(--) strength distribution for $^{82}$Se up to 30 MeV in Fig. \ref{fig13}.
Our running sum results for the $\beta^-$ strength, are 42 $\sim$ 54 for $\beta_2$= $\pm$ 0.1 $\sim$ $\pm$ 0.3.
The $\beta^-$ strength by the experimental data is 21.9 up to 12 MeV by Madey {\it et al.} \cite{Madey89}.
In Ref. \cite{Madey89}, they argued that the ISR, 3$(N-Z) = 52$, is satisfied about 52 \% if the $\beta^+$ strength is assumed to be zero.
One may notice that the running sum on $^{82}$Se calculated by $\beta_2$ = 0.3 represented by
blue dotted solid line seems to be most consistent with experimental data in Fig. \ref{fig13}.
The $\beta_2$ values for $^{82}$Se are 0.1934 (27) from the B(E2) and 0.133 from the RMF.

 In Table \ref{tab:XY}, we identify main collective $1^{+}$ states for the GT transitions on $^{82}$Se for $\beta_2$ = 0.3, which are presented as $X^{2} - Y^{2}$ of two dominant excitation energies having large B(GT) values, 3.88 and 2.77, located at $E_{ex}$ = 11.48 and 13.74 MeV, respectively, in Fig. \ref{fig11}(d). $X (Y)$ is a forward (backward) amplitude in Eq. (\ref{eq:tram_d}). The GT state at 11.48 MeV comes mostly from (422 3/2, 422 5/2) and (420 1/2, 431 3/2) configurations
and 13.74 MeV state comes from (413 5/2, 413 7/2), (310 1/2, 330 1/2), (202 3/2, 541 3/2) configurations. The particle states for the configuration are mainly located around the Fermi surface whose smearing becomes wider by the deformation \cite{sarriguren_a,Ha13_1}.

\subsubsection{$^{90}$Zr}
\begin{figure}
\includegraphics[width=0.8\linewidth]{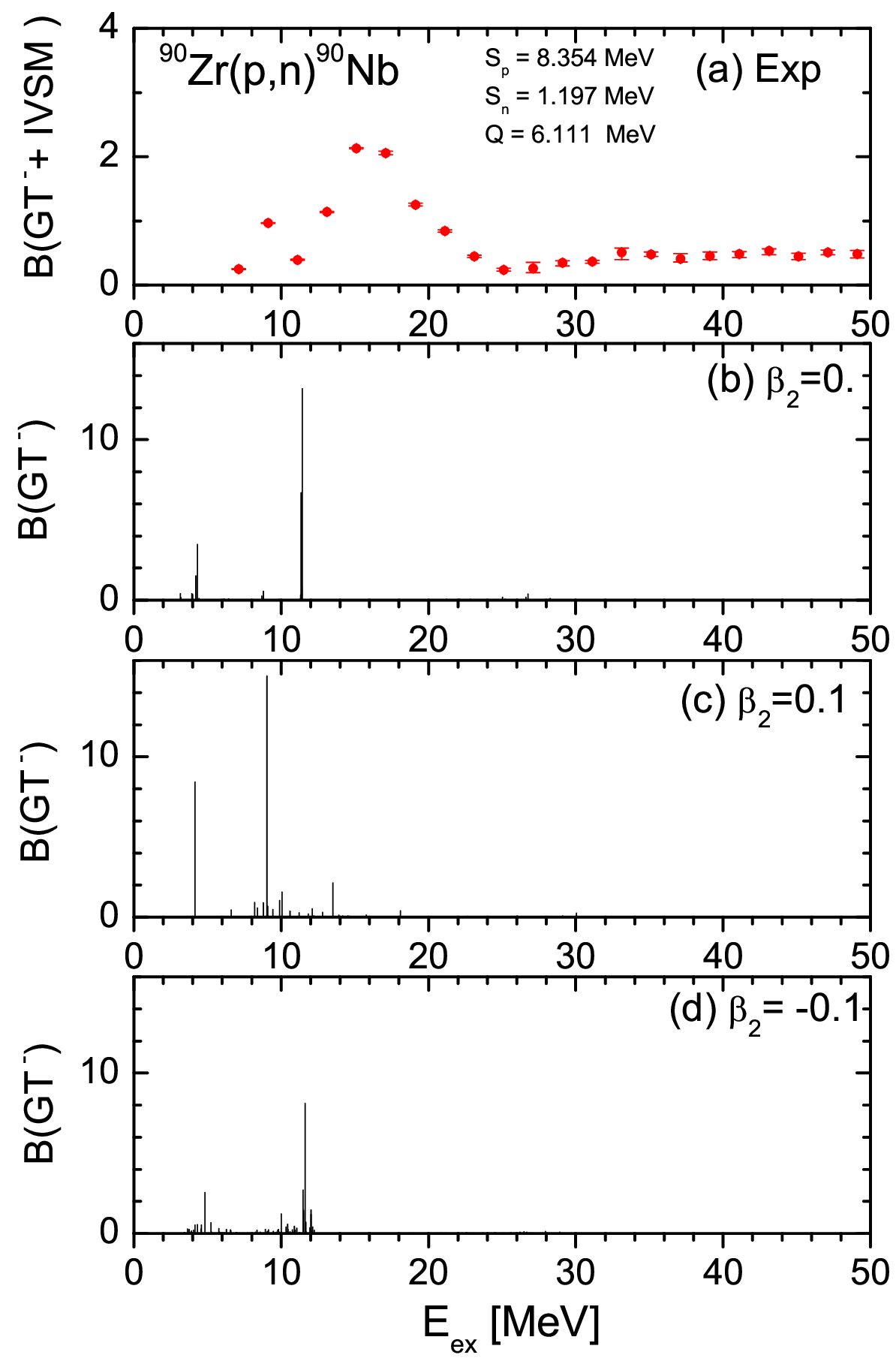}
\caption{(Color online) Gamow-Teller strength distributions B(GT$^{-}$)
on $ ^{90}$Zr as a function of the $E_{ex}$ w.r.t. the ground state of $^{90}$Zr. Experimental data denoted as filled
(red) points in the uppermost panel are deduced from the $^{90}$Zr(p,n) reaction at 293 MeV \cite{Yako05}. In each panel, we indicate
$\beta_2$. The IVSM excitation are thought to be located around 30 $\sim$ 38 MeV in panel (a).
$\beta_{2}^{RMF}$= 0.001 and $\beta_{2}^{E2}$= 0.089 (29) for $^{90}$Zr \cite{Lala99, Raman}.
}
\label{fig14}
\end{figure}

\begin{figure}
\includegraphics[width=0.8\linewidth]{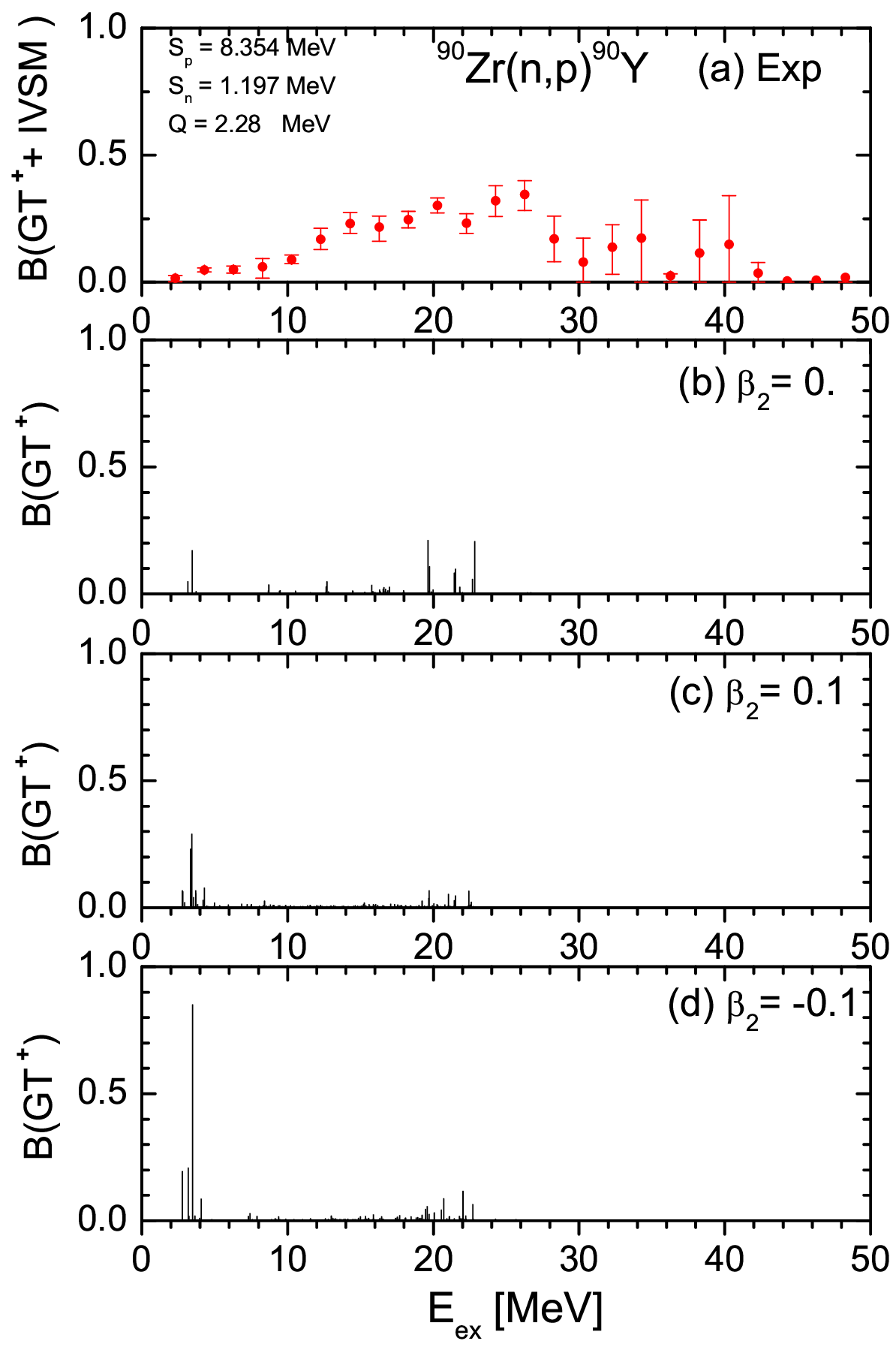}
\caption{(Color online) Gamow-Teller strength distributions B(GT$^{+}$)
on $ ^{90}$Zr as a function of the $E_{ex}$ w.r.t. the ground state of $^{90}$Zr.
The filled (red) circles in the uppermost panels are deduced from the $^{90}$Zr(n,p) reaction at 293 MeV \cite{Yako05}.
In each panel, we indicate $\beta_2$. The IVSM excitation in panel (a) are around 17 $\sim$ 25 MeV.}
\label{fig15}
\end{figure}

$^{90}$Zr is one of the most well known spherical nuclei. For instance, $\beta_2^{E2} = 0.089(29)$ and $\beta_2^{RMF} = 0.01$. Therefore it would be an interesting question if the DQRPA is a proper description of the excited states of spherical nuclei. Figs. \ref{fig14} and \ref{fig15} show the GT$(\mp)$ strength distributions on $^{90}$Zr as a
function of the $E_{ex}$ w.r.t. $^{90}$Zr, whose Q values for GT($\mp$) are 6.111 and 2.280 MeV, respectively. The uppermost panels (a) stand for the experimental data on GT(--)+IVSM (isovector spin monopole) and GT(+)+IVSM deduced from the $^{90}$Zr(p,n)$^{90}$Nb reaction $^{90}$Zr(n,p)$^{90}$Y at 293 MeV \cite{Yako05}, respectively. It should be noted that the IVSM contributions are included in the relatively higher excitation energy region because the IVSM transition operators are similar to the GT transition operators apart from additional radial factors \cite{Yako05}.

\begin{figure}
\includegraphics[width=0.8\linewidth]{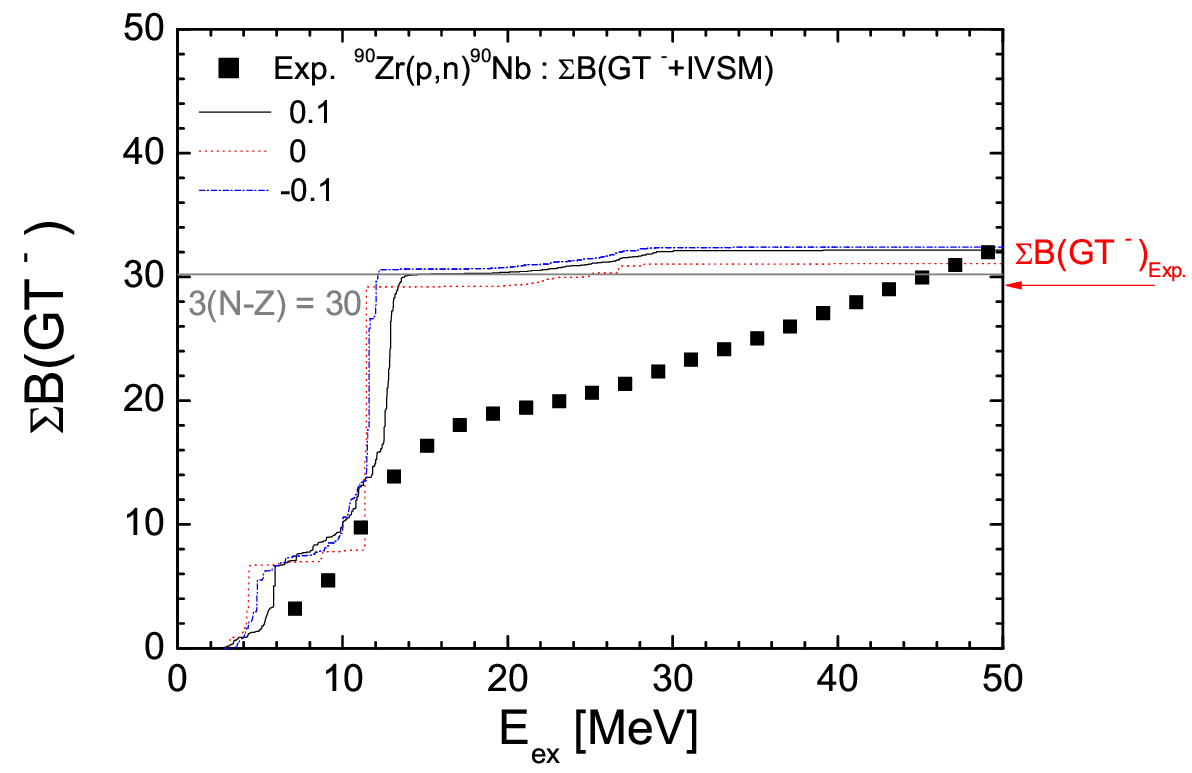}
\caption{(Color online) Running sums for the GT(--) strength distributions in Fig. \ref{fig14}.
Black squares are experimental data by Yako {\it et al.} from Ref. \cite{Yako05} which include
GT(-)+IVSM up to 50 MeV. Red arrow indicates the $\sum$B(GT$^{-}$) up to 50 MeV {\it i.e.} the contribution subtracted by the IVSM contribution.
}
\label{fig16}
\end{figure}

Panels (b)$\sim$(d) in Figs. \ref{fig14} and \ref{fig15} are the GT ($\mp$) strength results by the DQRPA for different $\beta_2$ values, where we did not include the IVSM because the contribution is known to be small compared to the GT contribution (see ${\Sigma}$ B(GT$^{-}$)$_{exp}$ in Fig. \ref{fig16}).
The ISR is almost satisfied at each $\beta_2$ value for both B(GT$(\mp)$) as shown in Table \ref{tab:parameters}.

However, our results of the GT strength distributions do not properly explain the data in the following two respects. First, peak positions of GT(--) are not recovered even if we take $\beta_2 = 0$ and $0.1$, which are thought to be more or less compatible with $\beta_2^{E2} = 0.089$, although the position of the GT excited energy around 14 MeV is deviated from the data only by 2 MeV for GT(--). Specifically, we do not have any noticeable high-lying GT states appeared in other nuclei.

\begin{figure}
\includegraphics[width=0.8\linewidth]{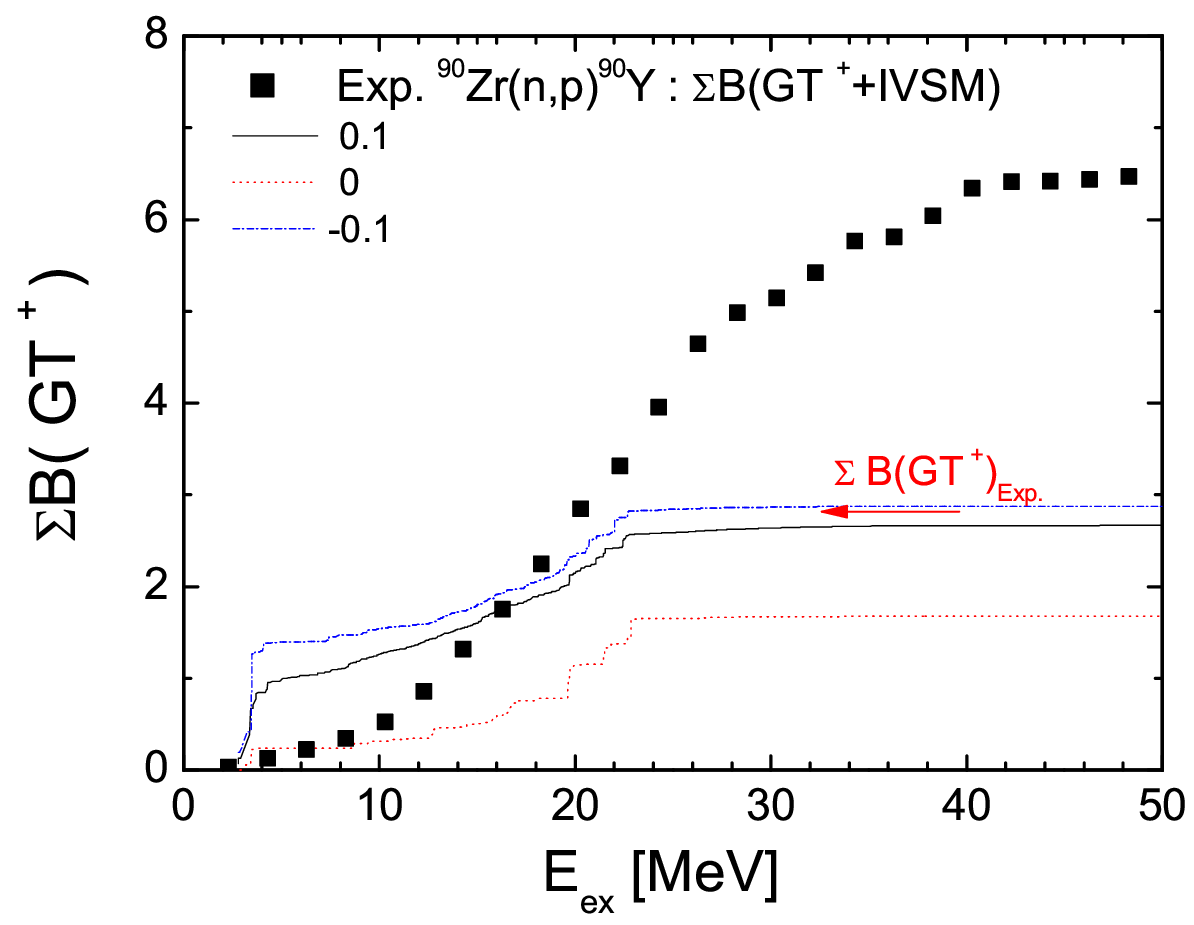}
\caption{(Color online) Running sums for the GT(+) strength distributions in Fig. \ref{fig15}.
Black squares are experimental data by Yako {\it et al.} from Ref. \cite{Yako05} which include
GT(+)+IVSM up to 32 MeV. Red arrow indicates the $\sum$B(GT$^{+}$) up to 32 MeV {\it i.e.} the contribution subtracted by the IVSM contribution.
}
\label{fig17}
\end{figure}

Second, running sums of the GT strength distribution shown in Figs. \ref{fig16} - \ref{fig17} show inconsistent results to the experimental data. Black squares (red arrow) are the accumulated sum of the GT ($\mp$) + IVSM (only of the GT ($\mp$)) strength data by Yako {\it et al.} deduced from Ref. \cite{Yako05}.
Since there are no GTGR peaks in the experimental data, the running sum show monotonous increase along with the $E_{ex}$, while theoretical results show a quantum leap by the GT strength around 5 MeV. Our results for the running sum are 37.29  and 7.91 in GT(-) and GT(+) at $\beta_2$=0.1 for DQRPA, while the running sum data of GT($\mp$)+IVSM (GT($\mp$)) are 33.01 (29.3) up to 50 MeV and
6.47 (2.9) up to 32 MeV. The ISR from Yako {\it et al.} is 88.47 $\%$ while the ISR calculated in the DQRPA is around 98 $\%$. One more interesting point in the data is that almost spherical nuclei, such as $^{90}$Zr, did not show any significant high-lying GT(--) excited states showed up on the $^{76}$Ge and $^{76,82}$Se data.

Here we argue that results of a spherical nucleus, $^{90}$Zr, are not as good as results of other deformed nuclei. First, if we recollect that the experimental data of GT$(\mp)$ strengths actually include the contributions by the IVSM \cite{Yako05} around 30 $\sim$ 38 MeV (17 $\sim$ 25 MeV) which are not considered in the present calculation, one may understand why our results do not show discernible high-lying GT strengths in Fig. \ref{fig16}. But, the high-lying states at the (GT(+)) transition is neatly explained as shown at (d) in Fig. \ref{fig15}.

Second, it arises from a problem inherent in the $\beta_2 = 0$ limit of the DQRPA. In fact, it is an interesting argument if our DQRPA formalism goes back to the spherical QRPA in the $\beta_2  = 0$ limit, or not. If we take the limit, a component of the deformed basis (the deformed harmonic oscillator wave function in Eq. (\ref{eq:trans})) goes back to a component of the spherical basis. Of course, for $\beta_2 \neq 0$ cases, one component of the deformed basis may have many components resulting from the spherical basis carrying the angular momentum higher than $j$ in $\Omega_j$, as shown in the following example
\begin{eqnarray}\label{eq:basis_d}
&& | {N n_z \Lambda \Sigma : 0 0 0 { 1 \over 2} } > = | 0s {1 \over 2} >  ~~;~\beta_{2} = 0 ~ , \\ \nonumber
     && | {N n_z \Lambda \Sigma : 0 0 0 { 1 \over 2} } > = 0.98 | 0s {1 \over 2} > + ~0.0005~ | 1s {1 \over 2} > + ~0.094 ~| 0d {5 \over 2} >  \\ \nonumber & & + ~ 0.0116~ | 0d {3 \over 2} > + \cdots
~~;~ \beta_{2} = 0.3~~.
\end{eqnarray}
However, single particle states in the deformed Woods-Saxon potential are linear combinations of the deformed basis, as shown in Eq. (\ref{eq:Nil}), stemming from the diagonalization of the total Hamiltonian in the Nilsson basis. Therefore, even if we take the $\beta_2 = 0$ limit, there remained still other contributions from other components in the deformed (Nilsson) basis which can be expanded into the spherical basis keeping the same projection $\Omega_j$. Namely, the coefficients $b^{\pm}_{N n_z \Omega_{\alpha}}$ in Eq. (\ref{eq:Nil}) have components more than one even for the $\beta_2 = 0$ limit. For example, the deformed s.p. state $| \Omega_j = {1 \over 2} >$ can be represented in terms of the deformed basis and finally expanded in terms of the spherical s. p. states for two $\beta_2 = 0$ and 0.1 cases
\begin{eqnarray}\label{eq:basis}
| {1 \over 2} > &=& 0.931| (N n_z \Lambda)~ 0 0 0 > + 0.208| 2 2 0 > - 0.294| 2 0 0 > + 0.018 | 4 4 0 > + \cdots ~  \\\nonumber
 &=& 0.93| 0s {1 \over 2} > -0.36| 1s {1 \over 2} > + 0.04| 2s {1 \over 2} > + 0.01 | 3s {1 \over 2} > + \cdots ~~;\beta_{2} = 0 , \\\nonumber
| {1 \over 2} > &=& 0.927| (N n_z \Lambda)~ 0 0 0 > + 0.224| 2 2 0 > - 0.296| 2 0 0 > + 0.024 | 4 4 0 > + \cdots ~ \\\nonumber
 &=& 0.92 | 0s {1 \over 2} > - 0.36 | 1s {1 \over 2} > -0.03 | 0d {3 \over 2} > + 0.04 | 0d {5 \over 2} > + \cdots
~~; \beta_{2} = 0.1~.
\end{eqnarray}
Here one may notice that, even for the $\beta_2 = 0$ limit, $| \Omega_j = { 1 \over 2} >$ state has other components coming from  $| n (\neq 0) s { 1 \over 2} >$ states although main component is $ | 0 s { 1 \over 2} >$. Therefore, the $\beta_2 = 0 $ limit in the DQRPA can not exactly go back to the spherical QRPA, but approximately go back to the spherical QRPA. Therefore we may conclude that our DQRPA is not appropriate to describe spherical nuclei. On the contrary, results by the spherical QRPA were shown to be more consistent to the experimental data as shown in our previous paper \cite{Ch12-1}.

\subsubsection{$^{92}$Zr}

\begin{figure}
\includegraphics[width=0.8\linewidth]{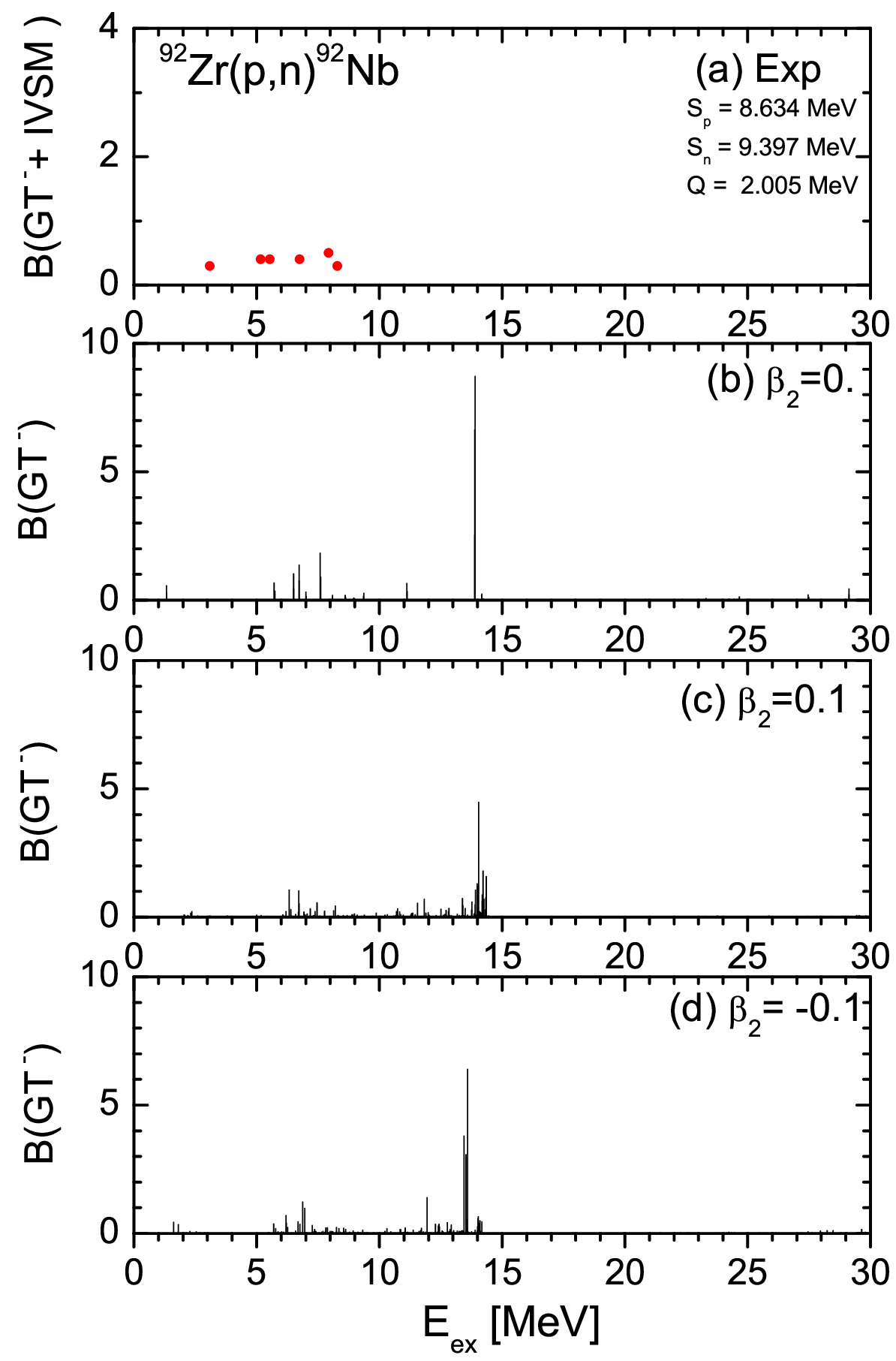}
\caption{(Color online) GT strength distributions B(GT$^{-}$)
on $ ^{92}$Zr as a function of the excitation energy $E_{ex}$ w.r.t. the ground state of $^{92}$Zr.
Experimental data denoted as filled (red) points in the uppermost panels are deduced from the $^{92}$Zr(p,n)$^{92}$Nb reaction at 26 MeV \cite{Bauer96}. In each panel, we indicate $\beta_2$.
$\beta_{2}^{RMF}$= 0.002 and $\beta_{2}^{E2}$= 0.103 (37) for $^{92}$Zr \cite{Lala99, Raman}.
}
\label{fig18}
\end{figure}

\begin{figure}
\includegraphics[width=0.8\linewidth]{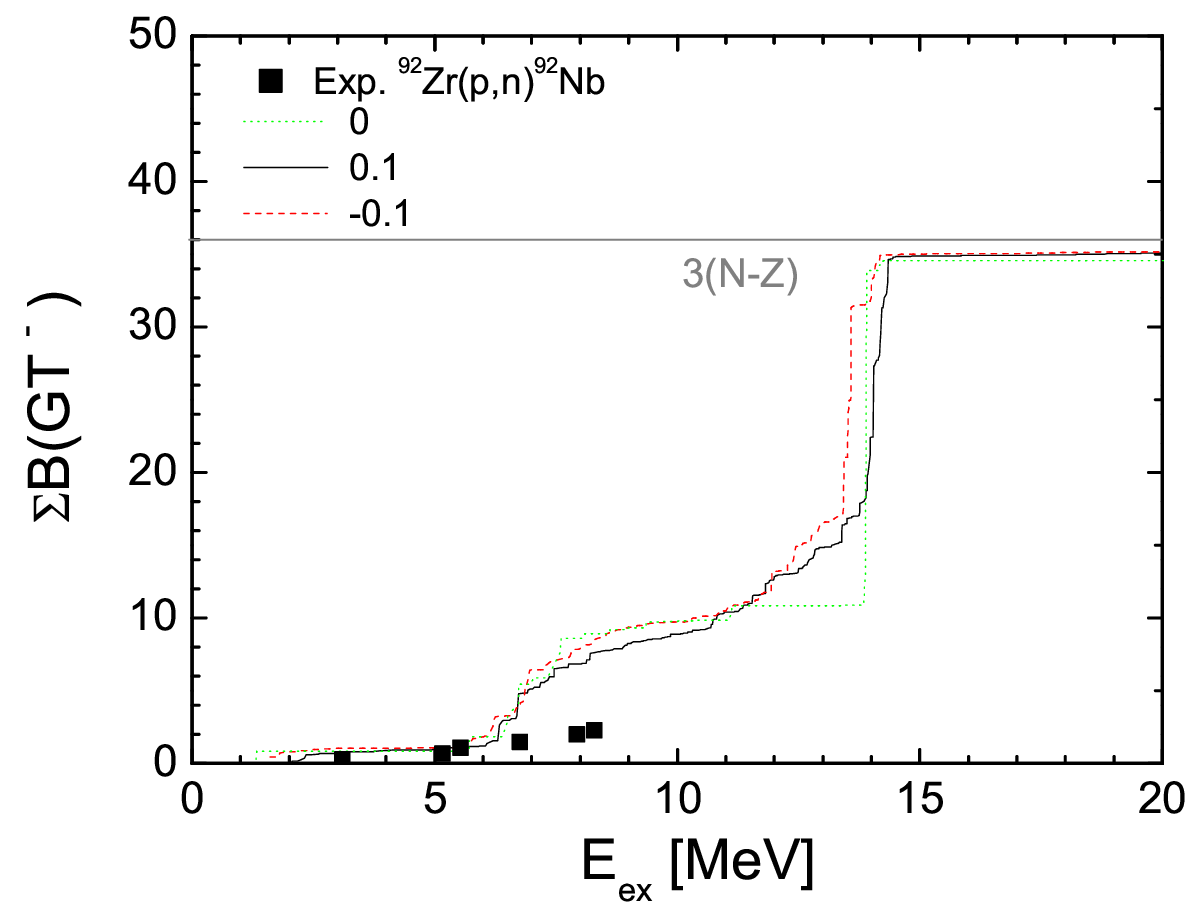}
\caption{(Color online) Running sums for the GT(--) strength distributions in Fig. \ref{fig18}. Experimental data are from the results by Grimes {\it et al.} at Ref. \cite{Bauer96}.}
\label{fig19}

\end{figure}

Figs. \ref{fig18} and \ref{fig19} show the GT(--) strength distributions and their running sum of $^{92}$Zr as a
function of the $E_{ex}$ w.r.t. the ground state of $^{92}$Zr, whose Q value is 2.005 MeV.
The uppermost panel (a) of Fig. \ref{fig18} represents the experimental B(GT$^{-}$) values extracted from $^{92}$Zr(p,n)$^{92}$Nb reaction at 26MeV. Since the projectile energy was too low to expect high-lying excited states, most GT excitation are observed below 9 MeV. Panels (b) $\sim$(d) are our results for $\beta_2$ = 0, 0.1, and -0.1. Results for $\beta_2 = \pm 0.1$, whose values are similar to $\beta_2^{E2} = 0.1027$, seem to be compatible with the experimental data. Our theoretical calculations address a possibility of another peak around 14 MeV at (c) and (d) in Fig. \ref{fig18}.

On the other hand, $^{92}$Zr(p, n)$^{92}$Nb reaction is very important to understand cosmological origin of $^{92}$Nb \cite{Haya13} which nucleus is thought to be produced mainly by the neutrino process. In particular, the neutrino-induced reaction via charged current, $^{92}$Zr($\nu_e, e^-$)$^{92}$Nb, was shown to be vital for the $^{92}$Nb production \cite{Ch12-2}. Therefore, more experimental data on $^{92}$Nb or $^{92}$Zr by charge conserving and/or exchanging reactions with higher energy projectiles are desirable for understanding of the origin of $^{92}$Nb as well as nuclear structures and reactions related to $^{92}$Nb and $^{92}$Zr.

\section{Summary and Conclusion}
To describe the nuclear structure and related nuclear reactions relevant to deformed nuclei, we developed the deformed QRPA which was progressed by exploiting a deformed axially symmetric Woods-Saxon potential, the
deformed BCS with a realistic two-body
interaction by Brueckner $G$-matrix based on Bonn
CD potential. Results of the Gamow-Teller strength by the DQRPA, for $^{76}$Ge, $^{76,82}$Se, and $^{90,92}$Zr show that the deformation effect leads to a fragmentation of the GT strength into high-lying GT excited states. In particular, in $^{76}$Ge and $^{82}$Se, which data were already measured at the charge exchange reaction experiments by the 134.4 MeV proton beam, the high-lying GT excited states beyond one nucleon threshold were found to be consistent with our calculations. Those high-lying GT excited states are shown to play a significant role of reproducing the Ikeda sum rule, which is believed to be satisfied irrespective of any nuclear models used for the deformation effects.

In order to confirm the high-lying GT states, we studied in detail their running sums for the GT($\pm$) strength distributions. Our results show that the running sum of the GT strength distributions for $^{76}$Ge and $^{76,82}$Se reproduce well experimental data without quenching factor. This fact is a remarkable point because one usually uses the quenching factor to satisfy the experimental GT strength data. But as shown in recent experimental results on the GT strength distributions on $^{90}$Zr, even the experimental data can satisfy the ISR sum rule up to 90\% by the addition of the high-lying GT strength data.

These high-lying excited GT states may affect seriously relevant nuclear reactions, particularly for neutrino-induced reactions exploited in the neutrino-process, because the emitted neutrinos from the proto-neutron star may have a high energy tail up to tens of MeV energy range. Since the experiments relate to the neutrino reaction on such a high energy would be a very challenging task in the present neutrino factories, the extraction of the high-lying GT states from various charge exchange reactions could be very useful for understanding the neutrino reaction in the cosmos, if we recollect that the GT transition is the main component for the neutrino-induced reaction.

We also tested the validity of our DQRPA by applying to $^{90}$Zr, which is believed to be almost spherical. It turns out that the present approach is not appropriate to apply to spherical nuclei, because the deformed single particle state $ |\Omega_j >$ by the deformed Woods-Saxon potential adopted for the DQRPA may have inevitable non-zero components in the deformed basis carrying total angular momentum higher than $j$, which may have projections $\Omega_j$, even if we take the deformation parameter $\beta_2 = 0$ limit.

More systematic analysis of deformed nuclei by our DQRPA are under progress for exotic light nuclei characterized by the shell evolution or the inversion island. In the light nuclei or neutron-deficient nuclei, which may have small energy gaps between protons and neutrons and small N-Z values, the neutron-proton pairing correlations could definitely affect the GT transition phenomena. Therefore more data for the GT($\mp$) strength distributions on the nuclei, in particular, on the high-lying GT states, could be a stringent test of deformed nuclear models.

Finally, since one of our calculations uses the expansion of a deformed basis into a spherical basis, any types of transitions (or operators) beyond the GT transition operator can be calculated straightforwardly in the spherical basis. More extensive calculations including electro-magnetic transitions as well as weak interaction transitions are in progress.

\section*{Acknowledgement}
We are grateful to Professors, Amand Faessler and Fedor Simkovic, for helpful discussions on this work. This work was supported by the National Research Foundation of
Korea (2012R1A1A3009733, 2012M7A1A2055605, and 2014R1A2A2A05003548).

\newpage
\vspace{2mm}
\begin{table}
\caption[bb]{Deformation parameters $\beta_2$ in Eq. (\ref{eq:beta}); $\beta_{2}^{E2}$ is the $\beta_2 $ value extracted from the E2 transition \cite{Raman} and
$\beta_{2}^{RMF}$ is the $\beta_2 $ value calculated from the RMF calculation \cite{Lala99}.
}
\setlength{\tabcolsep}{1.0 mm}
\begin{tabular}{cccccc}\hline
 $\beta_2$        & ${}^{76}$Ge &  ${}^{76}$Se   &  ${}^{82}$Se & ${}^{90}$Zr  & ${}^{92}$Zr  \\ \hline \hline
 $\beta_{2}^{E2}$ &  0.262(39)  &  0.309(37)     & 0.193(27)    &  0.089(29)   &  0.103(37)    \\ \hline
 $\beta_{2}^{RMF}$ &  0.157     &   -0.244       & 0.133        &   0.001      &  0.002 \\ \hline \hline
\end{tabular}
\label{tab:deformation}
\end{table}
\hspace{5cm}
\begin{table}
\caption[bb]{Deformation parameter $\beta_2$ exploited in Eq. (\ref{eq:Nil}), empirical
(theoretical) pairing gap parameters $\Delta^{\textrm{p,n}}_{\textrm{em}}$ ($\Delta^{\textrm{p,n}}_{\textrm{th}}$),
and proton (neutron) pairing strength parameters $g_{\textrm{pair}}^{p}(g_{\textrm{pair}}^{n})$ used in this work.
The ISR in the last column denotes the Ikeda sum rules I and II, Eqs. (\ref{eq:isr-d-I}) and (\ref{eq:isr-d-II}), as a percentage ratio to $3(N-Z)$, which show almost identical results. The particle-particle (particle-hole) strength parameters are exploited as $g_{\textrm{pp}}=0.99 ~ (g_{\textrm{ph}}=1.15)$ for all nuclei.}
\setlength{\tabcolsep}{1.0 mm}
\begin{tabular}{cccccccc}\hline
 nucleus & $\beta_2$  & $\Delta^{\textrm{p}}_{\textrm{em}}$(MeV)& $\Delta^{\textrm{p}}_{\textrm{th}}$(MeV)& $\Delta^{\textrm{n}}_{\textrm{em}}$(MeV)& $\Delta^{\textrm{n}}_{\textrm{th}}$(MeV) & $g_{\textrm{pair}}^{p}(g_{\textrm{pair}}^{n})$ & ISRI=II ($\%$)   \\ \hline \hline
            &  0.1 &              & 1.563    &        & 1.537  & 1.248(1.350)     & 98.77 \\
            &  0.2 &              & 1.562    &        & 1.536  & 1.344(1.320)     & 98.54 \\
${}^{76}$Ge &  0.3 &     1.562    & 1.563    & 1.535  & 1.535  & 1.545(1.406)     & 98.42 \\
            & -0.1 &              & 1.562    &        & 1.535  & 1.235(1.338)     & 98.77 \\
            & -0.2 &              & 1.562    &        & 1.535  & 1.244(1.398)     & 98.44 \\
            & -0.3 &              & 1.561    &        & 1.537  & 1.291(1.445)     & 98.38 \\ \hline
            &  0.1 &              & 1.755    &        & 1.709  & 1.279(1.420)     & 98.73 \\
            &  0.2 &              & 1.753    &        & 1.710  & 1.460(1.440)     & 98.74 \\
${}^{76}$Se &  0.3 &    1.753     & 1.755    & 1.709  & 1.711  & 1.667(1.439)     & 98.84 \\
            & -0.1 &              & 1.755    &        & 1.711  & 1.264(1.382)     & 98.74 \\
            & -0.2 &              & 1.755    &        & 1.709  & 1.231(1.402)     & 98.58 \\
            & -0.3 &              & 1.753    &        & 1.710  & 1.387(1.418)     & 98.64 \\ \hline
            &  0.1 &              & 1.411    &        & 1.711  & 1.136(1.463)     & 97.47 \\
            &  0.2 &              & 1.409    &        & 1.545  & 1.231(1.679)     & 97.65 \\
${}^{82}$Se &  0.3 &   1.409      & 1.410    & 1.544  & 1.546  & 1.484(1.663)     & 96.68 \\
            & -0.1 &              & 1.410    &        & 1.546  & 1.150(1.403)     & 98.00\\
            & -0.2 &              & 1.410    &        & 1.545  & 1.105(1.433)     & 97.76\\
            & -0.3 &              & 1.410    &        & 1.545  & 1.264(1.404)     & 97.93 \\\hline
            &  0.1 &              & 1.249    &        & 1.706  & 1.208(1.553)     & 98.27\\
${}^{90}$Zr &  0.  &   1.247      & 1.249    & 1.705  & 1.705  & 1.037(1.129)     & 97.92\\
            & -0.1 &              & 1.249    &        & 1.706  & 1.202(1.253)     & 98.42\\ \hline
            &  0.1 &              & 1.357    &        & 0.841  & 1.249(1.342)     & 97.96\\
${}^{92}$Zr &  0.  &   1.357      & 1.358    & 0.841  & 0.841  & 1.043(1.082)     & 97.96\\
            & -0.1 &              & 1.357    &        & 0.841  & 1.235(1.238)     & 98.05\\ \hline \hline
\end{tabular}
\label{tab:parameters}
\end{table}
\vspace{2mm}
\begin{table}
\caption[bb]{Configurations of two main GT states for $^{82}$Se in Fig. \ref{fig11}(d) presented by
$X^{2}-Y^{2}$, where $E_{ex}$ and $B(GT)$ stand for corresponding excited energy w.r.t. the parent nucleus and GT strength. Detailed explanations are in text.}
\setlength{\tabcolsep}{2.0 mm}
\begin{tabular}{cccc}\hline
 $E_{ex}(MeV)$ & $B(GT)$ & Configuration       & $ X^{2} - Y^{2} $  \\ \hline \hline
11.48          &  3.88   & (422 3/2, 422 5/2)  & 0.37 \\
               &         & (420 1/2, 431 3/2)  & 0.13 \\ \hline
13.74          &  2.77   & (413 5/2, 413 7/2)  & 0.30 \\
               &         & (310 1/2, 330 1/2)  & 0.21 \\
               &         & (202 3/2, 541 3/2)  & 0.13 \\ \hline \hline
\end{tabular}
\label{tab:XY}
\end{table}
\hspace{5cm}
\newpage


\end{document}